\documentclass[twocolumn]{aastex6}
\usepackage{graphicx}
\usepackage{amssymb,amsfonts,amsmath,amstext,amsgen,amsopn,amsxtra,indentfirst,times}

\usepackage[normalem]{ulem}

\newcommand{\real}{\mathbb{R}}

\begin{document}

\title{Interpreting high-resolution spectroscopy of exoplanets using \\ cross-correlations and supervised machine learning}

\author{Chloe Fisher\altaffilmark{1}}
\author{H. Jens Hoeijmakers\altaffilmark{1,3}}
\author{Daniel Kitzmann\altaffilmark{1}}
\author{Pablo M\'arquez-Neila\altaffilmark{2}}
\author{Simon L. Grimm\altaffilmark{1}}
\author{Raphael Sznitman\altaffilmark{2}}
\author{Kevin Heng\altaffilmark{1}}

\altaffiltext{1}{University of Bern, Center for Space and Habitability, Gesellschaftsstrasse 6, CH-3012, Bern, Switzerland; chloe.fisher@csh.unibe.ch, jens.hoeijmakers@space.unibe.ch, kevin.heng@csh.unibe.ch}
\altaffiltext{2}{ARTORG Center for Biomedical Engineering, University of Bern, Bern, Switzerland. }
\altaffiltext{3}{Observatoire astronomique de l'Universit\'{e} de Gen\`{e}ve, 51 Chemin des Maillettes, 1290 Versoix, Switzerland. }

\begin{abstract}
We present a new method for performing atmospheric retrieval on ground-based, high-resolution data of exoplanets. Our method combines cross-correlation functions with a random forest, a supervised machine learning technique, to overcome challenges associated with high-resolution data. A series of cross-correlation functions are concatenated to give a ``CCF-sequence" for each model atmosphere, which reduces the dimensionality by a factor of $\sim 100$. The random forest, trained on our grid of $\sim 65,000$ models, provides a likelihood-free method of retrieval. The pre-computed grid spans 31 values of both temperature and metallicity, and incorporates a realistic noise model. We apply our method to HARPS-N observations of the ultra-hot Jupiter KELT-9b, and obtain a metallicity consistent with solar ($\log{\rm M}=-0.2 \pm 0.2$). Our retrieved transit chord temperature ($T=6000^{+0}_{-200}$K) is unreliable as the ion cross-correlations lie outside of the training set, which we interpret as being indicative of missing physics in our atmospheric model. We compare our method to traditional nested-sampling, as well as other machine learning techniques, such as Bayesian neural networks. We demonstrate that the likelihood-free aspect of the random forest makes it more robust than nested-sampling to different error distributions, and that the Bayesian neural network we tested is unable to reproduce complex posteriors. We also address the claim in \cite{cobb19} that our random forest retrieval technique can be over-confident but incorrect. We show that this is an artefact of the training set, rather than the machine learning method, and that the posteriors agree with those obtained using nested-sampling.
\end{abstract}


\section{Introduction}
\label{sect:intro}

\subsection{Observational motivation I: the rise of ground-based high-resolution spectra}

\begin{table*}
\begin{center}
\caption{High-resolution cross-dispersed echelle (grating) spectrographs with wide instantaneous wavelength coverage.}
\label{tab:spectrographs}
\vspace{0.1in}
\begin{tabular}{lccccc}
\hline
\hline
Name & Telescope & Resolving power & Wavelength Range (nm) & Status & Reference(s) \\
\hline
HARPS & ESO 3.6 m& 120,000 & 378--691 & Active & \cite{mayor03} \\
HARPS-N & TNG & 120,000 & 378--691 & Active & \cite{cosentino12} \\
ESPRESSO & VLT & 70,000--190,000 & 378--691 & Active & \cite{pepe14} \\
CARMENES & CAHA 3.5 & 80,000--100,000 & 520--1710 & Active & \cite{quirrenbach10} \\
GIANO & TNG & 50,000 & 950--2450 & Active & \cite{origlia14}\\
CRIRES+ & VLT & 50,000--100,000 & Y, J, H, K, L, M bands & Under development & \cite{follert14} \\
UVES & VLT & 40,000--110,000 & 300--1100 & Active & \cite{dekker00} \\
NIRSPEC & Keck & 25,000 & 960 -- 5500 & Active & \cite{mclean98} \\
PEPSI & LBT & 43,000--270,000 & 383--912 &  Active & \cite{strassmeier15} \\
HDS & Subaru & 90,000--165,000 & 298--1016 &  Active & \cite{noguchi02}\\
EXPRES & DCT & 150,000 & 380--844 & Active & \cite{fischer17} \\
HIRES & ELT & 100,000 & 397--2500 & Under development & \cite{zerbi14} \\
NIRPS & ESO 3.6 m& 80,000 & 974--1809 & Under development & \cite{wildi17}\\
SPIRou & CFHT & 70,000 & 980--2440 & Active & \cite{donati18} \\
iShell & IRTF & 75,000 & J, H, K, L, M bands & Active & \cite{rayner16} \\
IGRINS & HJS & 40,000 & 1450--2450 & Active & \cite{park14} \\
\hline
\hline
\end{tabular}\\
\vspace{0.2in}
\end{center}
\end{table*}

The observational characterisation of exoplanetary atmospheres via the measurement of transmission and emission spectra is occurring on two fronts: low-resolution, space-based spectroscopy (mainly with the Hubble Space Telescope and Spitzer), and high-resolution spectroscopy using a wide variety of ground-based spectrographs (Table \ref{tab:spectrographs}).  Spectra measured from space have the advantage that the spectral continuum, which encodes information on chemistry and clouds/hazes, may be measured in an absolute sense.  Ground-based spectra lose the spectral continuum---and effectively measure \textit{relative} transit depths or fluxes---due to having to correct for the presence of the Earth's atmosphere, but offer the key advantage that individual spectral lines may be resolved with spectral resolution $\sim 10^5$.  A plausible approach is to combine the advantages each has to offer and jointly analyze space- and ground-based spectra (e.g., \citealt{brogi17}).

Following the pioneering work of \cite{snellen08,snellen10} (see also \citealt{wiedemann01,brown02,deming05}), the use of high-resolution, ground-based spectroscopy to identify the presence of atoms and molecules has become routine \citep{redfield08,brogi12,birkby13,birkby17,brogi13,brogi14,brogi18,dekok13,lockwood14,wyttenbach15,wyttenbach17,piskorz16,piskorz17,piskorz18,khalafinejad17,khalafinejad18,nugroho17,hoeijmakers18,hoeijmakers19,cauley19,guilluy19,seidel19}.  These identifications are essentially model independent, relying only on knowledge of the cross sections or opacities of these atoms and molecules as determined by quantum physics (e.g., \citealt{rothman98,heng17}).  Line transition databases contain the positions and relative strengths of individual lines, either from experimental measurement or derived from first principles, which are then cross-correlated against the lines detected in the high-resolution spectrum.  By matching dozens to hundreds of lines using cross-correlation, robust identifications of atoms and molecules may be obtained  (but see \citealt{hoeijmakers15,brogi19} for examples of detections being dependent on the accuracy of the line-database used to compute these opacities).  In contrast, the claimed detections of molecules other than water in the Wide Field Camera 3 (WFC3) spectra of exoplanetary atmospheres remains model-dependent and an active topic of debate (e.g., \citealt{fisher18}), because at these resolutions ($\sim 10$) only the shapes of the overall opacities, consisting of a large collection of lines averaged together, are measured.

Interpreting ground-based, high-resolution spectra using the cross-correlation technique has one major shortcoming: cross-correlation is mainly capable of answering the binary question of whether an atom or molecule is absent or present, either in emission or absorption.  It does not yield the abundance of that atom or molecule, nor the atmospheric temperature and pressure of the environment in which it lies.  It similarly does not yield cloud or haze properties of the atmosphere.  The first study to decisively address this shortcoming was \cite{brogi19}, who re-analyzed CRIRES observations and derived an analytical expression that maps the cross-correlation function to the likelihood function.  The ability to compute the likelihood function implies that Bayes's Theorem may subsequently be invoked to compute posterior distributions of chemical abundances, temperature, etc.

CRIRES was an infra-red echelle spectrograph mounted on UT1 of ESO's VLT \citep{kaufl04}. Although the spectrograph achieved high spectral resolution of $\sim100,000$, the instantaneous wavelength coverage was small because the spectrograph was not cross-dispersed. Consequently, the spectra analyzed by \cite{brogi19} contain only 4096 data points (1.9626--2.0045$\mu$m, 2.2875--2.3454$\mu$m in two different modes).  As every model being computed in the atmospheric retrieval needs to be cross-correlated against the spectrum, it becomes computationally prohibitive to scale this method up to spectra of cross-dispersed echelle spectrographs that contain $\sim 10^5$--$10^6$ data points, because this increases the computational time by a factor $\sim 10^2$--$10^3$.  However, elucidating such a scalable method is crucial in the era of high-resolution spectrographs with wide \textit{instantaneous} wavelength coverage, an overview of which we list in Table \ref{tab:spectrographs} (also see \cite{gibson20} for a retrieval on data from the blue arm of UVES using an MCMC method.)

A novel method to analyze ground-based, high-resolution spectra with $\sim 10^5$--$10^6$ data points is therefore needed that will allow the computational effort to be reduced at the order-of-magnitude level \textit{and} allow for the computation of posterior distributions of parameters.

\subsection{Observational motivation II: failure of direct retrievals on noisy spectra}

Another major limitation of ground-based, high-resolution spectra is the observational uncertainty. The level of noise on each individual spectral data point is typically much greater than the signal itself, which causes the direct retrieval to fail (see Section \ref{sec:failure_direct_retrieval}). While each individual spectral point contains little information, the entire spectrum does encode valuable information on the atmospheric abundances and properties. Any successful interpretation method needs to leverage the information content of the entire spectrum against the high level of noise present.  

This is the rationale behind the cross-correlation technique, which has been adopted by many workers (e.g., \citealt{snellen10,brogi12,birkby13,dekok13,lockwood14,wyttenbach15,piskorz16,nugroho17,hoeijmakers18,guilluy19,seidel19}), including \cite{brogi19}.

In the current study, we will incorporate the cross-correlation technique into a novel method for performing retrievals on noisy, high-resolution spectra, but in a way that is distinct from \cite{brogi19}.

\subsection{Theoretical motivation I: likelihood-free inference methods using machine learning}
\label{sec:intro_likelihood}

In the published exoplanet literature, atmospheric retrievals typically assume the likelihood function to be a Gaussian when implementing the Markov Chain Monte Carlo (MCMC) or nested-sampling routines (e.g., \citealt{benneke12,line13a,waldmann15,lavie17,macdonald17,fisher18,brogi19}),
\begin{equation}
\ln{\cal L} = -\frac{1}{2} \sum^n_i \left( \frac{R_i - R_{i,{\rm obs}}}{\sigma_i} \right)^2 - \frac{\ln{\left( 2 \pi \sigma_i^2 \right)}}{2},
\label{eq:likelihood}
\end{equation}
where the transmission spectrum has $n$ measurements of transit radii ($R_{i,{\rm obs}}$) that are compared to the theoretical values of the transit radii ($R_i$) computed using a model.  The standard deviation of the uncertainty on each data point, assumed to follow a Gaussian distribution, is $\sigma_i$.  It is further assumed that the uncertainties are uncorrelated with one another.

One of the motivations of the current study is to provide an alternative inference approach that is \textit{likelihood-free}, meaning that one does not have to explicitly assume the functional form of the likelihood function. In practice, these likelihood-free inference approaches belong to the class of Approximate Bayesian Computation (ABC) methods \citep{sisson19}. Specifically, we use the supervised machine learning method of the random forest \citep{ho98,breiman01}, which was previously adapted by \cite{marquez18} to interpret low-resolution Hubble-WFC3 transmission spectra.  The method relies on using a grid of pre-computed atmospheric models combined with an arbitrary noise model as a training set for the random forest.  The uncertainties on each data point in the measured spectrum are incorporated into the noise-free model grid to generate a training set of noisy models.  This approach is not unlike that of standard retrieval techniques, which typically compute a grid of atmospheric models on the fly.

The random forest consists of a collection of regression trees. Each regression tree is trained on a subset of the grid of atmospheric models. By identifying regions of the multi-dimensional parameter space that predict similar transmission spectra, each regression tree quantifies the ``distance" between the model and measured transmission spectra.  This plays the role of the Euclidean distance ($R_i - R_{i,{\rm obs}}$) in the Gaussian likelihood function, except that the likelihood is implicitly learned from the training set of noisy models. (See Section \ref{sec:random_forest} for more information about the random forest).

Other advantages offered by the random forest retrieval method include the ability to run large suites of mock retrievals to both validate the model grid used and quantify its sensitivity to the parameters, as well as information content analysis to quantify the relative importance of each data point in the spectrum towards determining the value of each parameter \citep{marquez18}.

\subsection{Theoretical motivation II: feature engineering}

Feature engineering is the process by which the training set used in a machine learning method is optimised, e.g., a reduction in the dimensionality of the problem.  Deep learning methods perform feature engineering in an automated way, but they are significantly more expensive to implement than the random forest.  One of the novel aspects of the current study is the use of feature engineering to efficiently interpret noisy, high-resolution spectra.  Instead of using the spectra themselves as the training set, we demonstrate that it is sufficient to use a set of cross-correlation functions (CCFs) that sparsely sample the parameter space.  The resulting ``cross-correlation sequence" serves as the training set for the random forest, resulting in a reduction in the size of the training set by a factor $\sim 100$.  This feature engineering step allows the random forest retrieval method to be scaled up to interpret high-resolution spectra with $\sim 10^5$--$10^6$ data points in a computationally feasible way.

\subsection{Layout of study}

In Section \ref{sec:methods}, we describe our methodology, including the computation of the model grid of transmission spectra (radiative transfer, opacities, chemistry), the implementation of the random forest method, etc.  In Section \ref{sec:results}, we show our results from testing the method, and also the retrieval on HARPS-N observations of KELT-9b. In Section \ref{sec:discussion}, we discuss the results and compare our method to nested-sampling and other machine learning techniques. In Section \ref{sec:conclusion}, we summarise our conclusions. 

\section{Methods}
\label{sec:methods}

\subsection{KELT-9b}

\begin{figure*}
\begin{center}
\includegraphics[width=2\columnwidth]{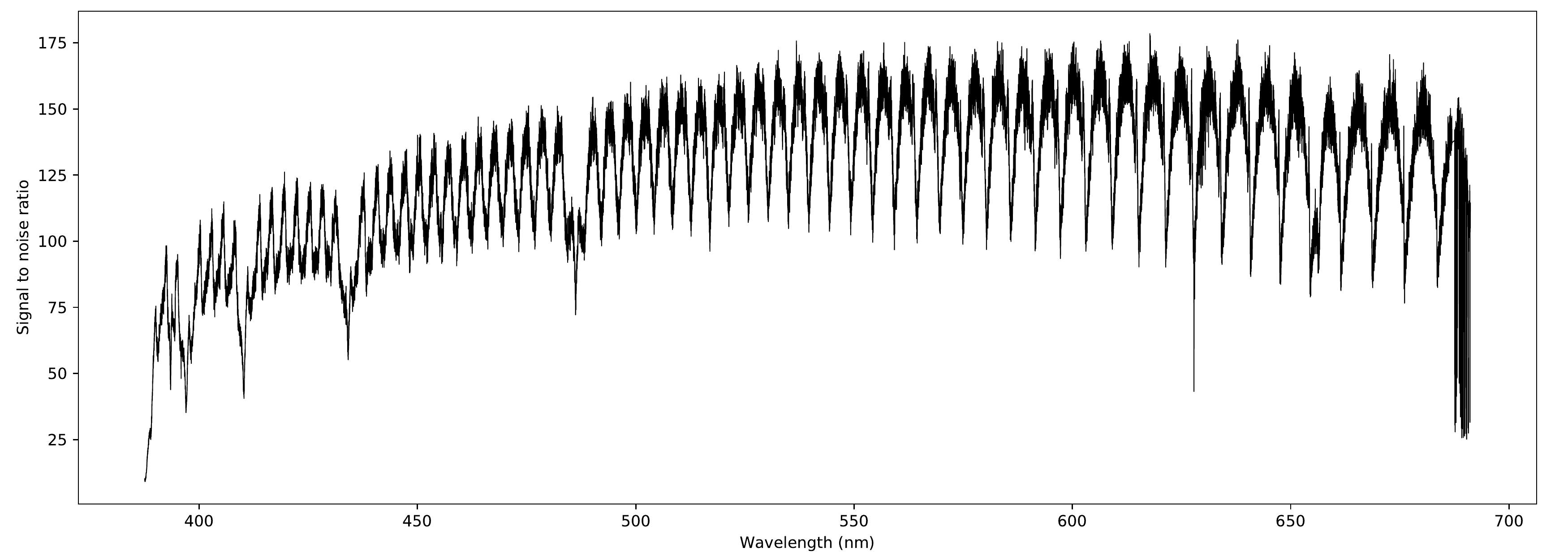}
\end{center}
\caption{The signal-to-noise level of the spectrum of the host star KELT-9 achieved in a 600 s exposure obtained with the HARPS-N instrument. The signal-to-noise is dominated by the photon (shot) noise, which decreases towards shorter wavelengths due to a reduced efficiency of the instrument, transmission of the Earth's atmosphere and lower intrinsic luminosity of the star. The significant narrow-band variation is due to the efficiency of the spectrograph falling off at the edges of spectral orders, as well as absorption lines in the star and the Earth's atmosphere.}
\label{fig:KELT9_snr}
\end{figure*}

As a proof of concept and in order to test the method, we have focused the retrieval on the ultra-hot Jupiter, KELT-9b. The brightness of the star combined with the extremely high temperatures allow for a higher signal to noise ratio than for other exoplanets (see Figure \ref{fig:KELT9_snr}), making it a good test subject for a retrieval on ground-based data. Furthermore, this object has been previously studied with high-resolution data in \cite{hoeijmakers18,hoeijmakers19}. \cite{kitzmann18} demonstrated that chemical equilibrium is a reasonable assumption, significantly reducing the number of parameters required in the atmospheric model, and that it is cloud-free with a continuum dominated by H$^{-}$ \citep{arcangeli18}. However, \cite{hoeijmakers19} suggested that there is most likely missing physics in this model, due to the discrepancy between the expected cross-correlation function for Fe$^{+}$ and the one obtained from the data. We will discuss this further in Section \ref{sec:kelt9b_retrieval}.

\subsection{Model Grid}
\label{sec:model_grid}

To construct the grid of models of KELT-9b, we adopt the system parameters reported by \cite{gaudi17} and \cite{hoeijmakers19}. We generate the models using an observation simulator, \texttt{Helios-o} \citep{bower19}, which follows the method described in \cite{gaidos17}. This algorithm has been validated in \cite{hengkitzmann17}, where it was compared against the models from \cite{fortney10}, \cite{deming13} and \cite{line13b}.

The model atmosphere is one-dimensional, plane-parallel, isothermal, in hydrostatic equilibrium and in chemical equilibrium.  It has 199 layers with 200 pressure levels ranging from $10^{-15}$--2 bar.  Each one-dimensional model atmosphere may be visualized as an atmospheric column.  Ray tracing is performed through a collection of these atmospheric columns to construct the transit chord at each wavelength,  taking into account the variation of gravity as different pressure levels are probed.  The variation of the effective transit radius with wavelength due to the chemical composition of the atmosphere is the transmission spectrum \citep{brown01}.  

The volume mixing ratios (relative abundances by number) of atoms, ions and molecules are computed using the \texttt{FastChem} chemical-equilibrium code, which considers gas-phase chemistry for more than 550 molecular species with elements more abundant than germanium \citep{stock18}. Additionally, we add most of the firstly and doubly ionized ions as well as anions for atoms lighter than neptunium \citep{hoeijmakers19}.  Our volume mixing ratios computed using \texttt{FastChem} are pressure-dependent, because of our non-isobaric treatment of the transit chord.  The opacities are computed using the open-source \texttt{HELIOS-K} opacity calculator \citep{grimm15}.  The inputs for the Fe, Fe$^+$, Ti and Ti$^+$ opacities are sourced from the Kurucz database\footnote{http://kurucz.harvard.edu/} \citep{kurucz17}.  The hydrogen anion (H$^-$) cross section is taken from \cite{john88}.  For completeness, collision-induced absorption associated with H-He, H$_2$-H$_2$ and H$_2$-He collisions are included \citep{richard12}.  Pressure broadening is neglected as the spectral continuum in ultra-hot Jupiters is dominated by absorption associated with the hydrogen anion (H$^-$), which masks the line wings.  The line shape is assumed to be a Voigt profile. The natural line width and thermal broadening are included \citep{kurucz17}.  Opacities are sampled uniformly across wavenumber with a spectral resolution of 0.01 cm$^{-1}$, and the transmission spectra are calculated at a resolution of 0.03 cm$^{-1}$.

The assumption of chemical equilibrium allows us to greatly simplify the theoretical analysis because the abundances of atoms and ions are completely specified by the temperature, pressure and elemental abundances. By assuming the ratios of elemental abundances follow those of the Sun, we reduce the chemical parameters down to a single number known as the metallicity. Therefore, we have just two parameters in our model --- temperature and metallicity. The temperature range of the grid spans from 3000 to 6000 K, in steps of 100 K, and the metallicity ranges from 0.1 to 100 times solar (-1 to 2 for the logarithm of the metallicity, $\log{\rm M}$, in steps of 0.1). This results in 31 values for each parameter, and thus 961 models in the grid in total. 

\subsection{Modeling HARPS-N observations}
We use existing observations of KELT-9b produced by the HARPS-N spectrograph \citep{hoeijmakers18} to convert the resulting model grid to models of the observed transmission spectrum. First, the transmission spectrum is convolved with a Gaussian with a full-width-at-half-maximum of $2.7$ km s$^{-1}$ (equivalent to the resolving power of the HARPS-N spectrograph), as well as a rotation-broadening profile that matches the rotation period of KELT-9b. It is subsequently interpolated onto the wavelength grid of the stitched, resampled pipeline-reduced (s1d) observations from HARPS-N. The continuum of the transmission spectrum is removed using a high-pass filter, in the same way as the observations with the HARPS-N spectrograph are filtered to remove broad-band spectral variations that are due to the instrument and variable observing conditions \citep{hoeijmakers18}. 

It would be possible to use this retrieval method for other instruments, such as those listed in Table \ref{tab:spectrographs}, however these would require different training sets to account for other observational effects. The noise model (see Section \ref{sec:noise_model}) would also need to be adjusted for different instruments.

\subsection{CCF-Sequences}
\label{sec:ccf_sequences}

We use the cross-correlation operator defined as
\begin{equation}
C(v) = \frac{\sum_i F_i {\cal T}_i(v)}{\sum_i {\cal T}_i(v)} ,
\end{equation}
where $F$ is the transmission spectrum, ${\cal T}$ is the cross-correlation template interpolated onto the same wavelength grid as the spectrum, $v$ is the velocity, and the summation takes place over the spectral data points. The denominator is a normalization factor, and thus the fluxes of the templates do not need to be rescaled when performing the cross-correlation.

Four subsets of cross-correlation templates, consisting of the spectral lines of neutral iron (Fe), singly-ionized iron (Fe$^+$), neutral titanium (Ti) and singly-ionized titanium (Ti$^+$), are created. Within each subset, there are 16 templates consisting of 4 values of temperature ($3000, 4000, 5000, 6000$ K) and 4 values of metallicity ($0.1, 1, 10, 100~\times$ solar).  In total, there are 64 cross-correlation templates. These templates are generated in the same way as the models (Section \ref{sec:model_grid}) with all but the relevant species' opacities removed from the final model, leaving only the required species' spectral lines. Broadening is not included as we are not aiming to retrieve dynamic properties. (See Section \ref{sec:vv_space} for tests involving velocity parameters.)

\begin{figure*}
\begin{center}
\hspace*{-0.8cm}
\includegraphics[width=2.4\columnwidth]{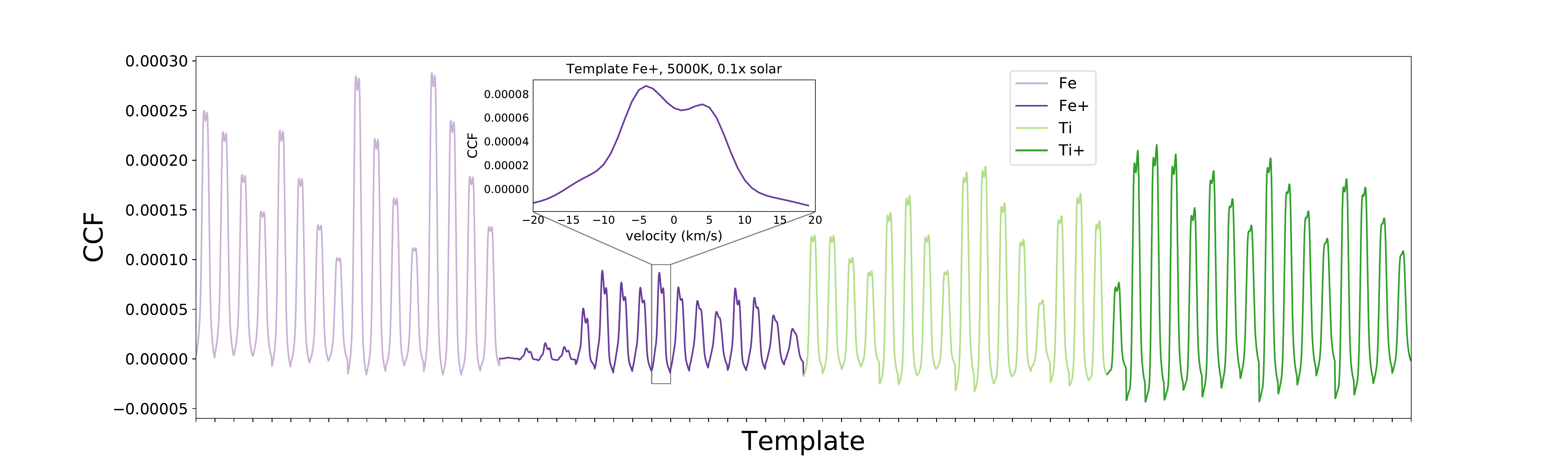}
\end{center}
\caption{Example of a CCF-sequence constructed by cross-correlating 64 templates with a model transmission spectrum with $T=3500$ K and $\log{\rm M}=0.8$. Each CCF has 40 points across velocity for a total of 2560 points for the entire CCF-sequence.  The insert magnifies one of the CCFs (Fe$^+$, $T=5000$ K and $\log{\rm M}=0.1$) for illustration.}
\label{fig:ccf_seq}
\end{figure*}

Each synthetic transmission spectrum in the model grid is cross-correlated with each of the 64 templates to create a set of 64 cross-correlation functions (CCFs). Additionally, each template is shifted in velocity space from -20 km s$^{-1}$ to 20 km s$^{-1}$ in steps of 1 km s$^{-1}$, resulting in 40 CCF values per template. These 64 CCFs are concatenated together to give a single sequence containing 2560 points, which we term a ``CCF-sequence" (Figure \ref{fig:ccf_seq}). Each of the 64 templates probes different components of the information contained in the spectral lines. In this way, the resulting CCF-sequence encodes the physical properties of the atmosphere over multiple axes. This feature engineering step has essentially reduced the dimensions of the model spectra by a factor of $\sim 100$.

\subsection{Noise Model}
\label{sec:noise_model}

Because KELT-9 is a bright star, the noise is dominated by photon-noise, and the SNR mainly varies due to the wavelength-dependent efficiency of the instrument, the stellar spectrum and Earth's atmospheric transmission function (see Figure \ref{fig:KELT9_snr}). The noise per spectral pixel is empirically measured from the time-series of observations used by \citet{hoeijmakers18}. For each spectral pixel a value may be drawn randomly from an assumed Gaussian distribution, creating a model of the noise of the entire spectrum that can be propagated through the cross-correlation function.

We assume each point in the spectrum $F$ has a Gaussian error bar with standard deviation $\sigma_{F_i}$. The noise model for the CCF then becomes a linear combination of Gaussians, therefore also a Gaussian, with a variance of
\begin{equation}
{\sigma^2_C} = \frac{\sum_i {\sigma^2_{F_i} }{{\cal T}_i(v)}^2}{\sum_i {{\cal T}_i(v)}^2} .
\label{eq:variance}
\end{equation}

We can then add the noise to the model grid of CCF-sequences. Since we require many instances of noise for the random forest, and the cross-correlation is computationally quite expensive, this provides a great advantage over applying the CCF to the noisy spectra.

\subsection{Random Forest}

\subsubsection{Theory}
\label{sec:random_forest}

The random forest consists of a collection of regression trees -- decision trees for interpreting continuous data. Each regression tree is trained on a subset of the grid of atmospheric models. During training, a tree is constructed by locating divisions in each wavelength dimension that sort the training spectra into groups with similar parameter values, known as leaves. Each leaf then has an assigned set of parameter values given by the training spectra in its group. When predicting on a real dataset, the spectrum is passed down each tree until it lands in a leaf, and the predicted parameter values are given by the corresponding set. The sets for every tree in the forest are then combined to give a distribution for each parameter.

The random forest falls into a class of inference methods known as ``Approximate Bayesian computation" (ABC; \citealt{sisson19}). ABC methods were invented to treat problems where it was either infeasible or impossible to explicitly specify the functional form of the likelihood (e.g., in the study of human populations).  Instead of seeking the maximum likelihood in a multi-dimensional parameter space, ABC methods seek to minimise some abstract distance (with the Euclidean distance being one specific example) between a set of simulated models and data to below some stated tolerance (Chapter 1.3 of \citealt{sisson19}).  If the tolerance is formally zero, then ABC methods become exact Bayesian methods, which have been shown to produce accurate posteriors (Chapter 1.6 of \citealt{sisson19}).  In practice, non-zero tolerances generally imply that the computed posterior distributions are approximate (hence the ``A" in ``ABC"), where the degree of accuracy depends on the tolerance specified (Chapter 1.5 of \citealt{sisson19}).  ABC methods often employ ``summary statistics" as a dimensionality reduction step (Chapter 1.7 of \citealt{sisson19}).  In the current study, the use of the CCF-sequence qualifies as a use of summary statistics.

\subsubsection{Setup}
\label{sec:setup}

\begin{figure}
\begin{center}
\includegraphics[width=\columnwidth]{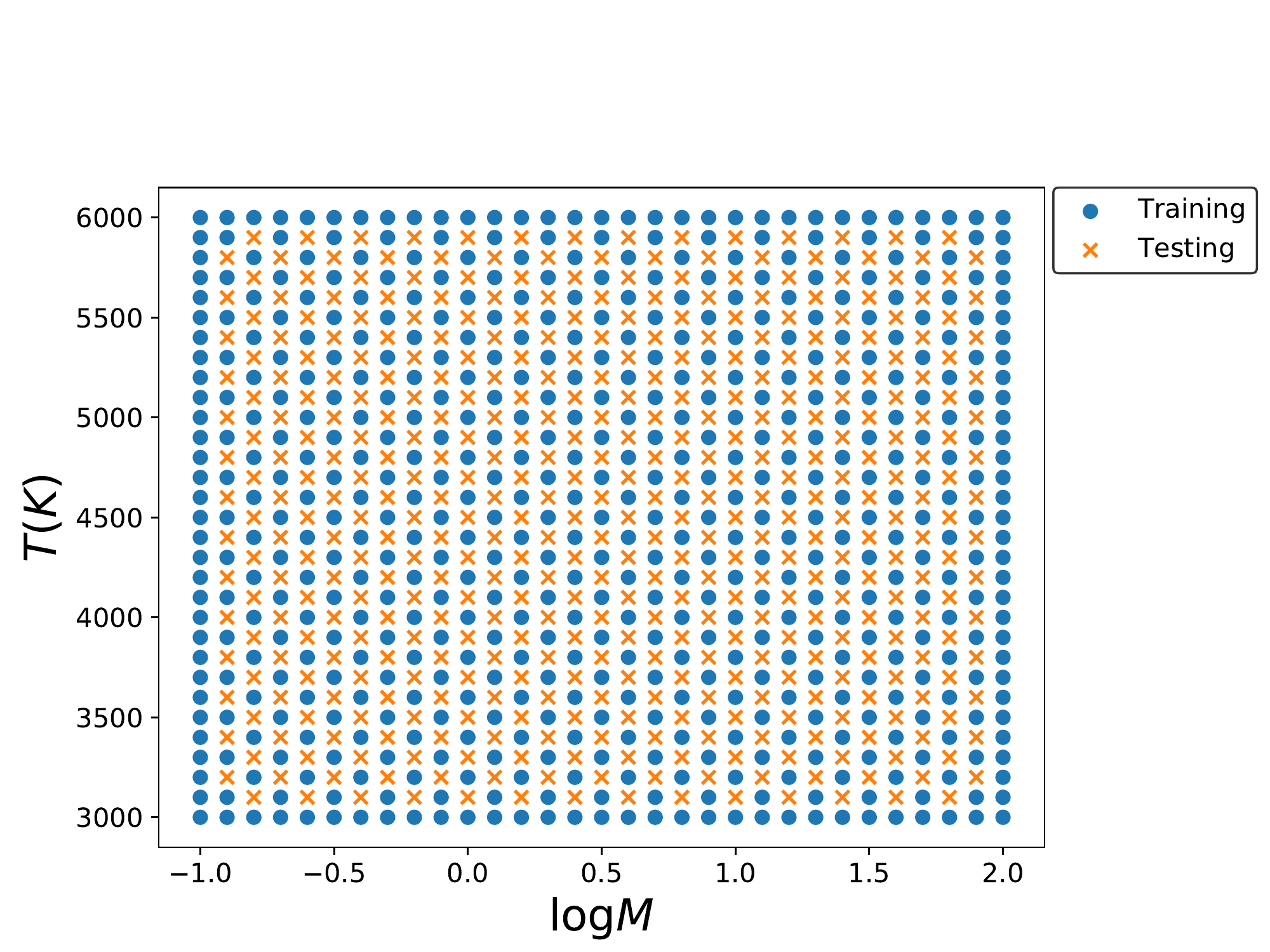}
\end{center}
\caption{Separation of the 961 members of the model grid into training and testing sets for the random forest.  The edges of this parameter space are intentionally included in the training set as the forest is unable to extrapolate.}
\label{fig:train_test_params}
\end{figure}

Starting from our grid of CCF-sequences, we divide the parameter space into training and testing sets, as shown in Figure \ref{fig:train_test_params}. This is to ensure the two sets are sufficiently distinct such that we can accurately test the performance of the forest. Next, we sample each point of the CCF-sequence within its respective uncertainty to generate 120 noisy instances of each CCF-sequence. We do this by drawing from Gaussian distributions with variance defined by equation \ref{eq:variance}. The entire set therefore amounts to 115,320 noisy CCF-sequences, with 64,920 in training and 50,400 in testing.

Our random forests consists of 1000 trees. Tree splitting is performed using a threshold variance of 0.01. Each time a tree is split, a random subset of 50 (approximately the square root) of the 2560 sequence points is used. Tree pruning methods are not used (see \citealt{breiman84,hastie01} for clarification of the terminology). For the predictions, the data is passed down through each tree until it reaches an end point, known as a leaf. The set of all training parameters that lie in this leaf are then given as the prediction for that tree. We call this the ``full-leaf" prediction. These training parameters come from the bootstrapped training dataset ---built using random sampling with replacement from the original training dataset--- that was used to train each tree. The final posterior is constructed by combining these predictions for all of the 1000 trees. This full-leaf prediction is an improvement on the previous method in \cite{marquez18}, in which only the mean parameter values corresponding to the predicted leaf were used, as it gives a more accurate approximation of the posterior. 

The implementation of the random forest method and $R^2$ metric are adopted from the open-source \texttt{scikit.learn} library (\citealt{pedregosa11}) in the \texttt{Python} programming language.  

\section{Results}
\label{sec:results}

\subsection{Failure of Direct Retrieval}
\label{sec:failure_direct_retrieval}

\begin{figure*}[t]
\centering
\begin{minipage}{.47\textwidth}
  \centering
  \includegraphics[width=.93\columnwidth]{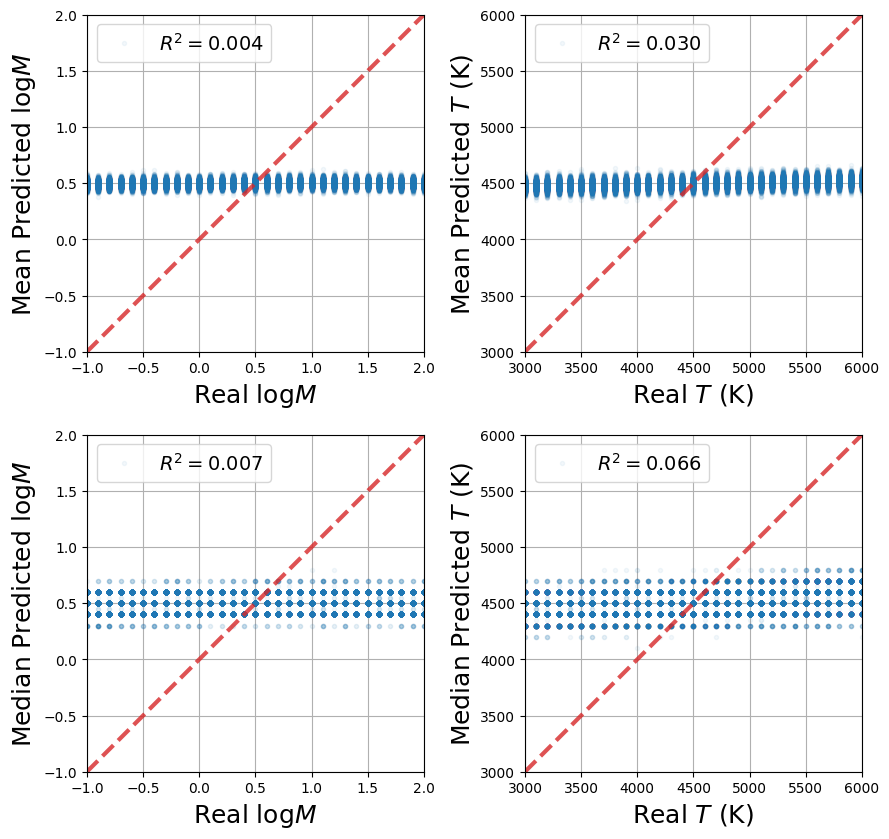}
  \caption{Predicted vs real values of the logarithm of metallicity ($\log{\rm M}$) and temperature ($T$) for the random forest trained using a section of the high-resolution spectrum containing $10^4$ points from 400 to 410 nm. The top and bottom sets of plots correspond to the mean and median predictions, respectively. The coefficient of determination ($R^2$) varies from -1 to 1, where values near unity indicate strong anti-correlations or correlations between the real and predicted values of a given parameter, based on the variance of outcomes. See Figure \ref{fig:failed_sec_retrieval} for a mock retrieval.}
  \label{fig:failed_sec_predvreal}
\end{minipage}\hfill%
\begin{minipage}{.47\textwidth}
  \centering
  \includegraphics[width=\columnwidth]{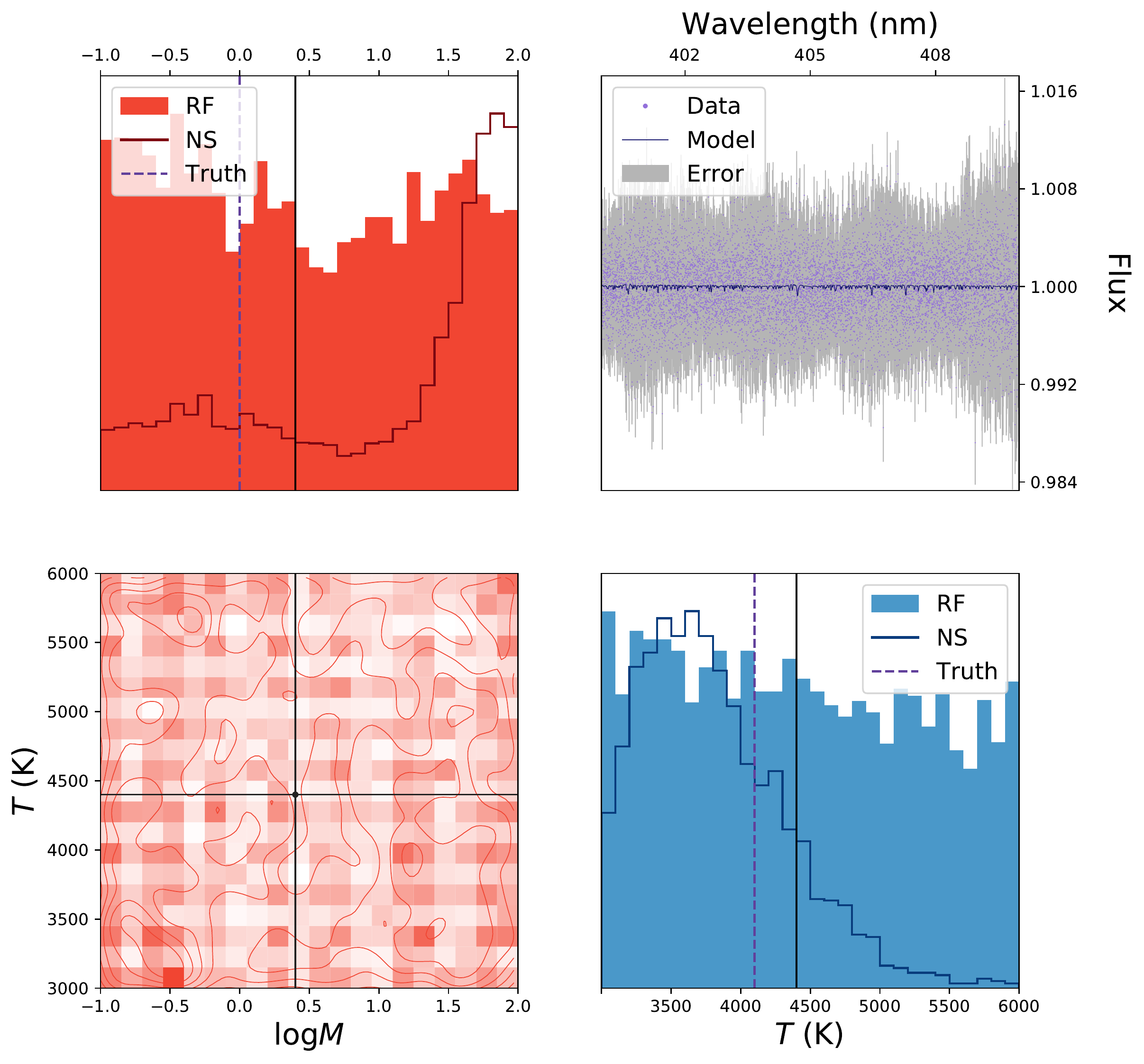}
  \caption{A mock retrieval using a section of the high-resolution spectrum containing $10^4$ points from 400 to 410 nm, from the test set shown in Figure \ref{fig:failed_sec_predvreal}. The mock spectrum has solar metallicity and a temperature of 4100 K. In the top left and bottom right panels, the solid posteriors show the results of the retrieval using the random forest (RF), and the empty line posteriors show the results from nested-sampling (NS). The purple, dashed lines show the true values. The top right panel shows the data points (lilac) with the error region (grey), along with the model (dark purple) corresponding to the medians from the $\log{M}$ and $T$ posteriors.}
  \label{fig:failed_sec_retrieval}
\end{minipage}
\end{figure*}

Initially we attempted to perform the random forest retrieval directly on the transmission spectra, set up in the same way as described in Section \ref{sec:setup} but with the model spectra instead of the CCF-sequences. Since the random forest method has been demonstrated to work for a dimensionality of at most $\sim10^4$ \citep{hastie01,sznitman13,zikic14,rieke15,zhang17}, we consider only a section of $10^4$ wavelength points from 400 to 410 nm in each synthetic spectrum.  Other sampling strategies (e.g., selecting line peaks only) produce similar outcomes\footnote{Whilst selecting line peaks is conceptually similar to a cross-correlation, by not averaging the points the noise remains high and hence the retrieval still fails.} (not shown). Figure \ref{fig:failed_sec_predvreal} shows the results of testing this forest, using both the mean (top panels) and median (bottom panels) predictions. The coefficient of determination, $R^2$, which measures the degree of agreement between the real versus predicted parameter values, is essentially zero for temperature and metallicity for both mean and median predictions, implying that the random forest has no predictive power when applied to the synthetic spectra themselves. Figure \ref{fig:failed_sec_retrieval} includes an example of the posterior distributions of temperature and metallicity for a mock retrieval, which are unconstrained and consistent with their prior distributions. In addition, we tested a traditional retrieval algorithm using nested-sampling \citep{skilling06,feroz08,feroz09,feroz13} with the open-source \texttt{PyMultinest} package \citep{buchner14}. Due to the high number of spectral points and complex forward model, we are unable to compute models on the fly as in a regular nested-sampling retrieval (see Section \ref{sec:nested-sampling}). Instead, we take the same grid of models as the forest, but without the added noise, and interpolate on it to produce forward models. Figure \ref{fig:failed_sec_retrieval} also shows the results from the nested-sampling mock retrieval. These posteriors span essentially the whole prior, with peaks offset from the correct values.

In summary, ground-based high-resolution spectra of exoplanets reside in a qualitatively different regime than the same measurements of stars or space-based low- to medium-resolution spectra of exoplanets. Individual data points hold little information as they are overwhelmed by noise, but the entire spectrum does encode useful information. This motivates our use of the cross-correlation functions, which effectively select the most informative lines in the spectrum.

\subsection{Random Forest Mock Retrievals}
\label{sec:rf_mocks}

\begin{figure*}[t]
\centering
\begin{minipage}{.47\textwidth}
  \centering
  \includegraphics[width=.93\columnwidth]{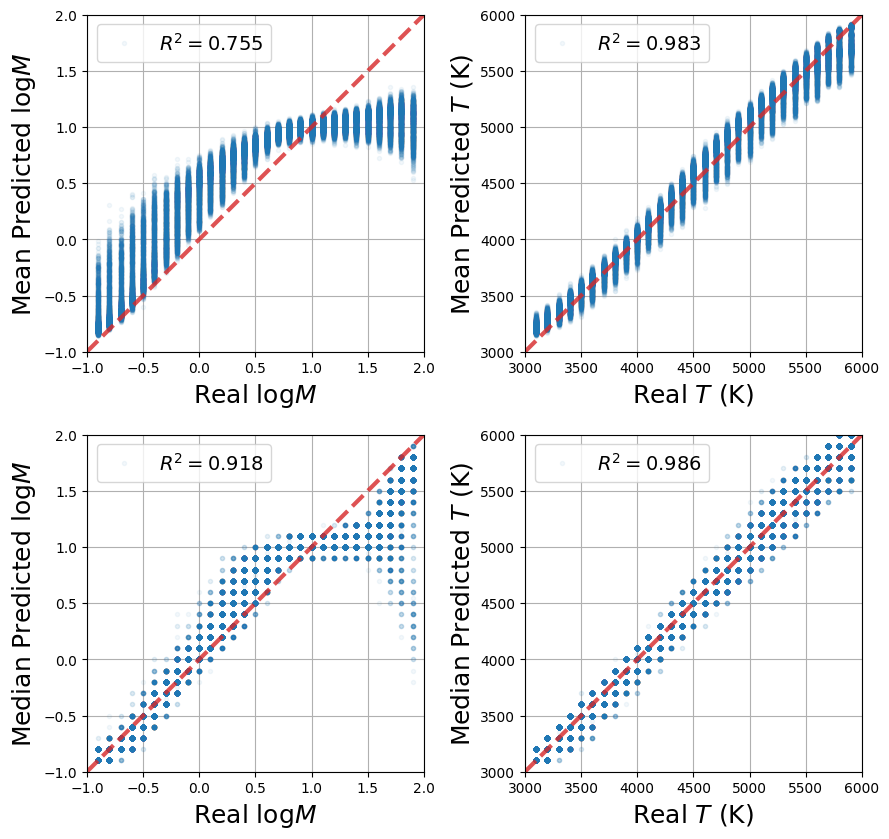}
  \caption{Predicted vs real values of the logarithm of metallicity ($\log{\rm M}$) and temperature ($T$) for the random forest trained on the CCF-sequences. The top and bottom panels show the results using the mean and median predictions, respectively. The coefficient of determination ($R^2$) varies from -1 to 1, where values near unity indicate strong anti-correlations or correlations between the real and predicted values of a given parameter, based on the variance of the outcomes. See Figure \ref{fig:ccf_mock} for a mock retrieval.}
  \label{fig:ccf_predvreal}
\end{minipage}\hfill%
\begin{minipage}{.47\textwidth}
  \centering
  \vspace*{-0.5cm}
 \includegraphics[width=\columnwidth]{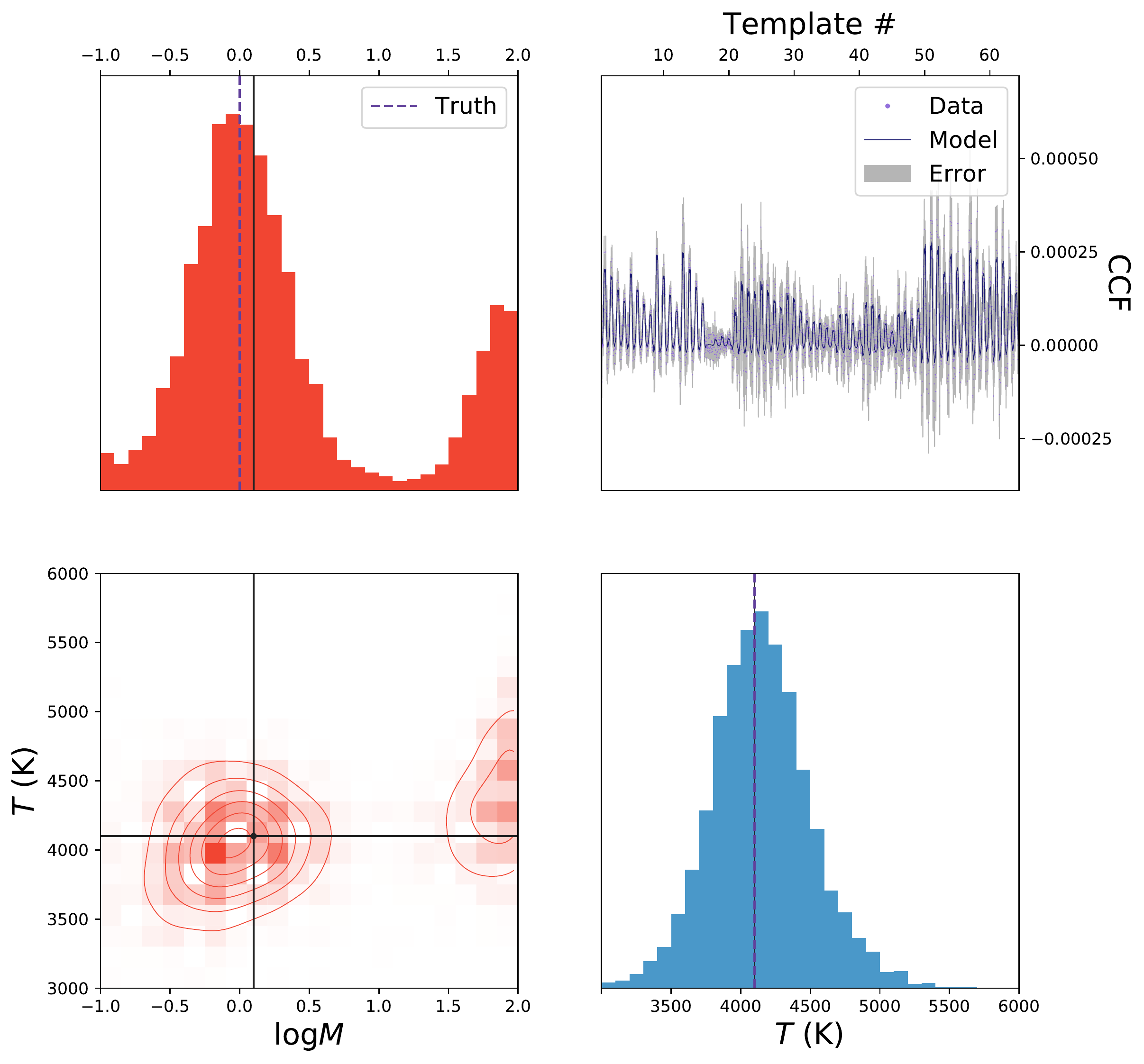}
  \caption{A mock retrieval performed on a model with solar metallicity and $T=4100$K, using the random forest trained on the CCF-sequences (see Figure \ref{fig:ccf_predvreal}). The black lines show the median values. The purple, dashed lines show the true values. The top right panel shows the data points (lilac) with the error region (grey), along with the model (dark purple) corresponding to the medians from the $\log{M}$ and $T$ posteriors.}
  \label{fig:ccf_mock}
\end{minipage}
\end{figure*}

Figure \ref{fig:ccf_predvreal} shows the results of testing the random forest trained on the CCF-sequences. The predictive power of the random forest has increased significantly. The difference in the predictability of the two parameters, metallicity and temperature, follows our intuition. The strength of spectral features are proportional to the logarithm of the opacity multiplied by the abundance of an atom. Because opacities have an exponential dependence on temperature \citep{rothman98,heng17}, the line strengths are highly sensitive to temperature and the ability of the random forest to predict temperature is strong. The ability to predict metallicity is somewhat weaker, because the metallicity linearly controls the atomic abundances, the logarithm of which determines the line-depths (e.g, \citealt{hengkitzmann17}). At high metallicities, the predictive power of the random forest tapers off, because the pressure scale height of the atmosphere decreases and the size of spectral features starts to decrease (see Section \ref{sec:metallicity_degeneracy}). The top and bottom panels of Figure \ref{fig:ccf_predvreal} correspond to the mean and median predictions of the trees, respectively. Traditionally, random forests produce mean predictions, but given the focus of atmospheric retrieval on posteriors and confidence intervals, we are more interested in the medians, which are more robust against asymetric posteriors. The increase in $R^2$ scores when using the median comes particularly from these more complex posteriors. Figure \ref{fig:ccf_mock} also shows an example of the posterior distributions obtained from the hybrid CCF retrieval, which recovers the injected values of temperature and metallicity accurately.

\begin{figure*}
\begin{center}
\hspace*{-0.5cm}
\includegraphics[width=2.0\columnwidth]{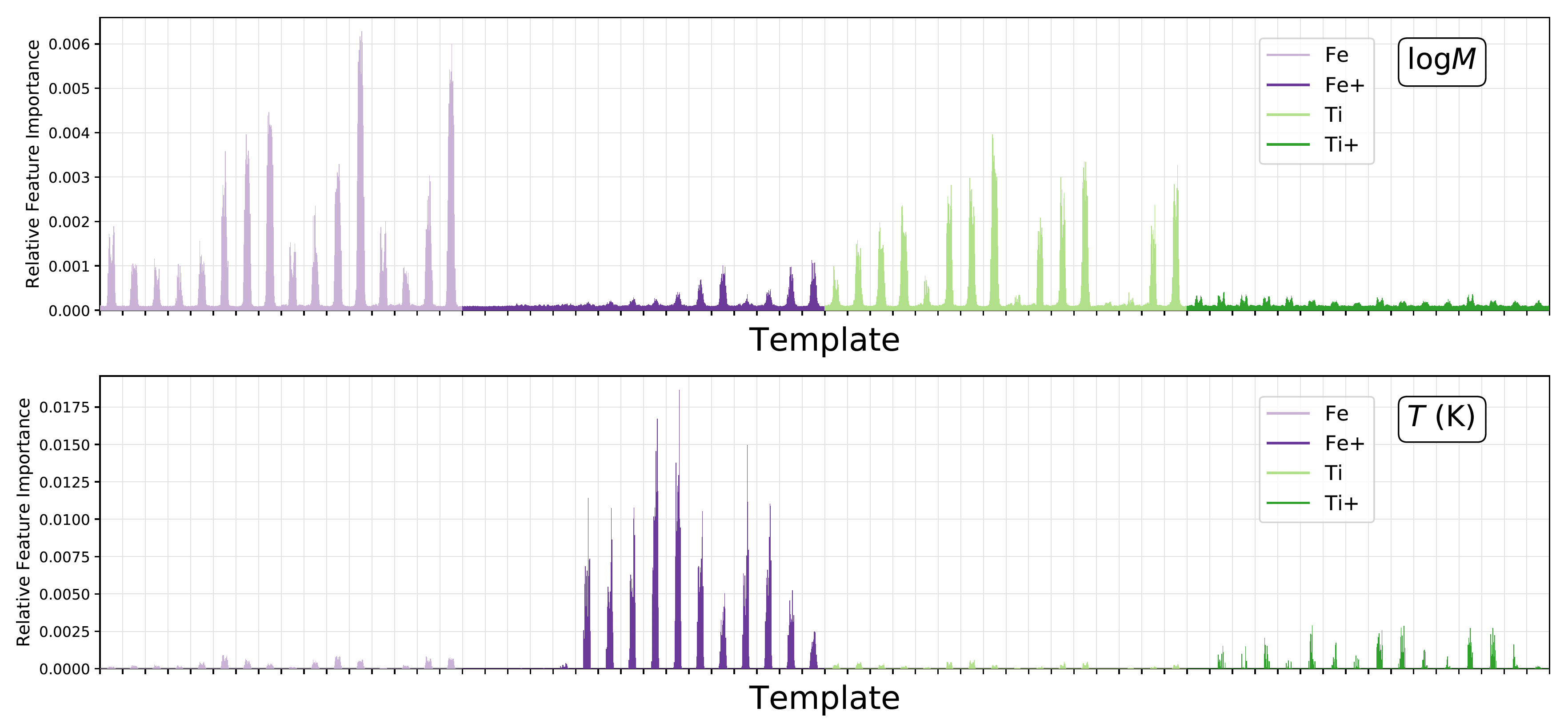}
\end{center}
\caption{Feature importance plots describing the relative importance of each CCF in the sequence towards constraining metallicity and temperature.}
\label{fig:feat_imp}
\end{figure*}

A useful, natural outcome of the random forest is the information content analysis known as the ``feature importance". This determines which data points hold the most importance for retrieving each parameter. Figure \ref{fig:feat_imp} shows the feature importance when predicting metallicity and temperature. As suggested by the bottom panel of Figure \ref{fig:feat_imp}, the ion species control the temperature prediction. Rising temperatures cause the neutral species to collisionally ionise, initially increasing the abundances of Fe$^+$ and Ti$^+$ by orders of magnitude while the corresponding decrease in neutral abundance is relatively small. 

As the metallicity increases, the depths of all metal absorption lines will tend to increase. However, in Figure \ref{fig:feat_imp} there appears to be a greater feature importance for the neutrals when predicting metallicity. A possible explanation for this is that as metallicity increases, the atmosphere will be more laden by free electrons from easily ionised species. Following the Saha equation \citep{saha20}, this will lead to a decrease in the ionisation fraction, partially negating the enhancement to the ion mixing-ratios that stems from the higher metal abundance. Therefore, the neutral species are more sensitive to metallicity.

\subsection{Metallicity Degeneracy}
\label{sec:metallicity_degeneracy}

\begin{figure}
\begin{center}
\includegraphics[width=\columnwidth]{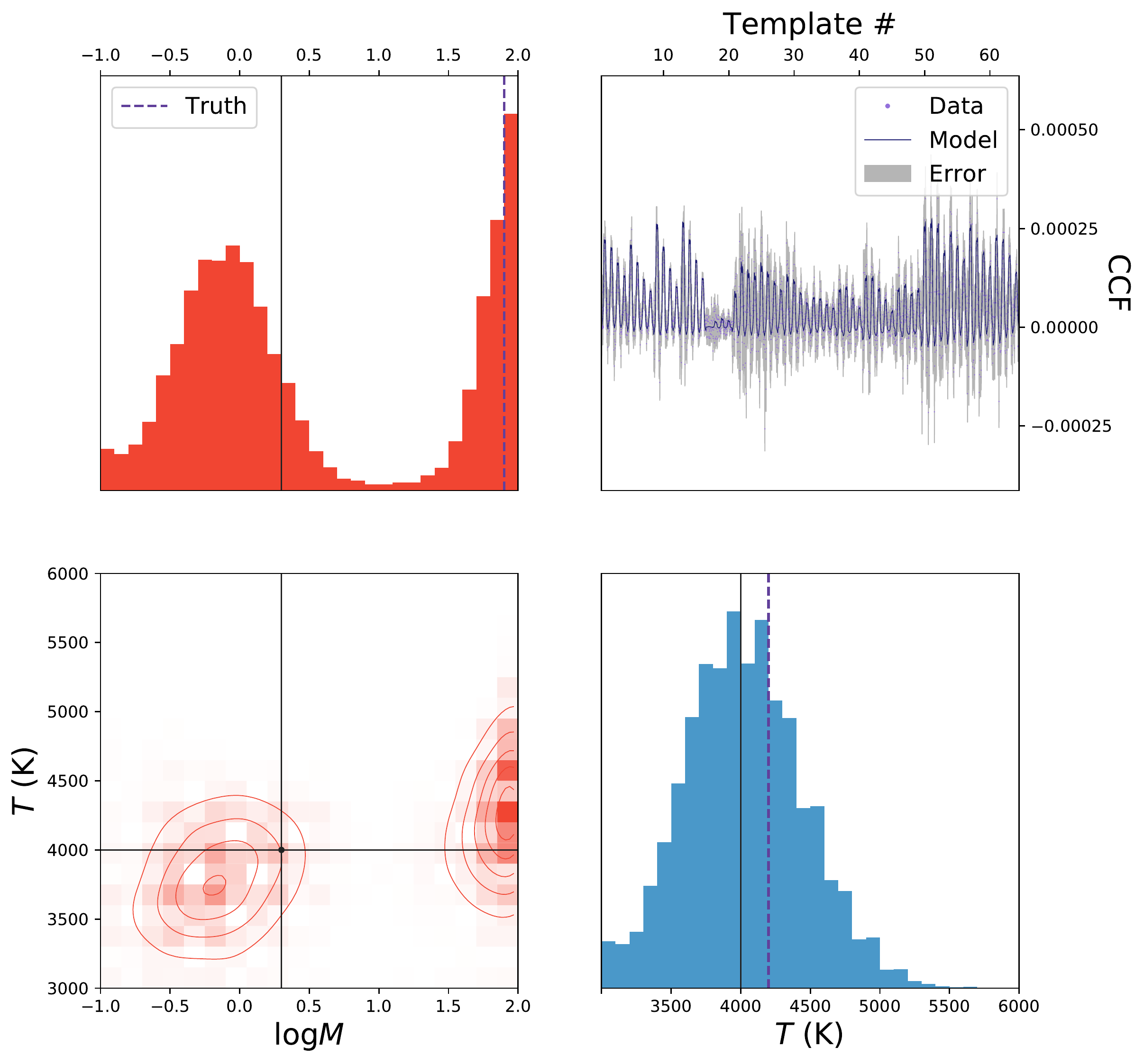}
\end{center}
\caption{A mock retrieval performed on a model with $\log{\rm M}=1.9$ and $T=4200$K, using the random forest trained on the CCF-sequences (see Figure \ref{fig:ccf_predvreal}). The black lines show the median values. The purple, dashed lines show the true values. The top right panel shows the data points (lilac) with the error region (grey), along with the model (dark purple) corresponding to the medians from the $\log{M}$ and $T$ posteriors.}
\label{fig:met_degen_mock}
\end{figure}

\begin{figure}
\begin{center}
\includegraphics[width=\columnwidth]{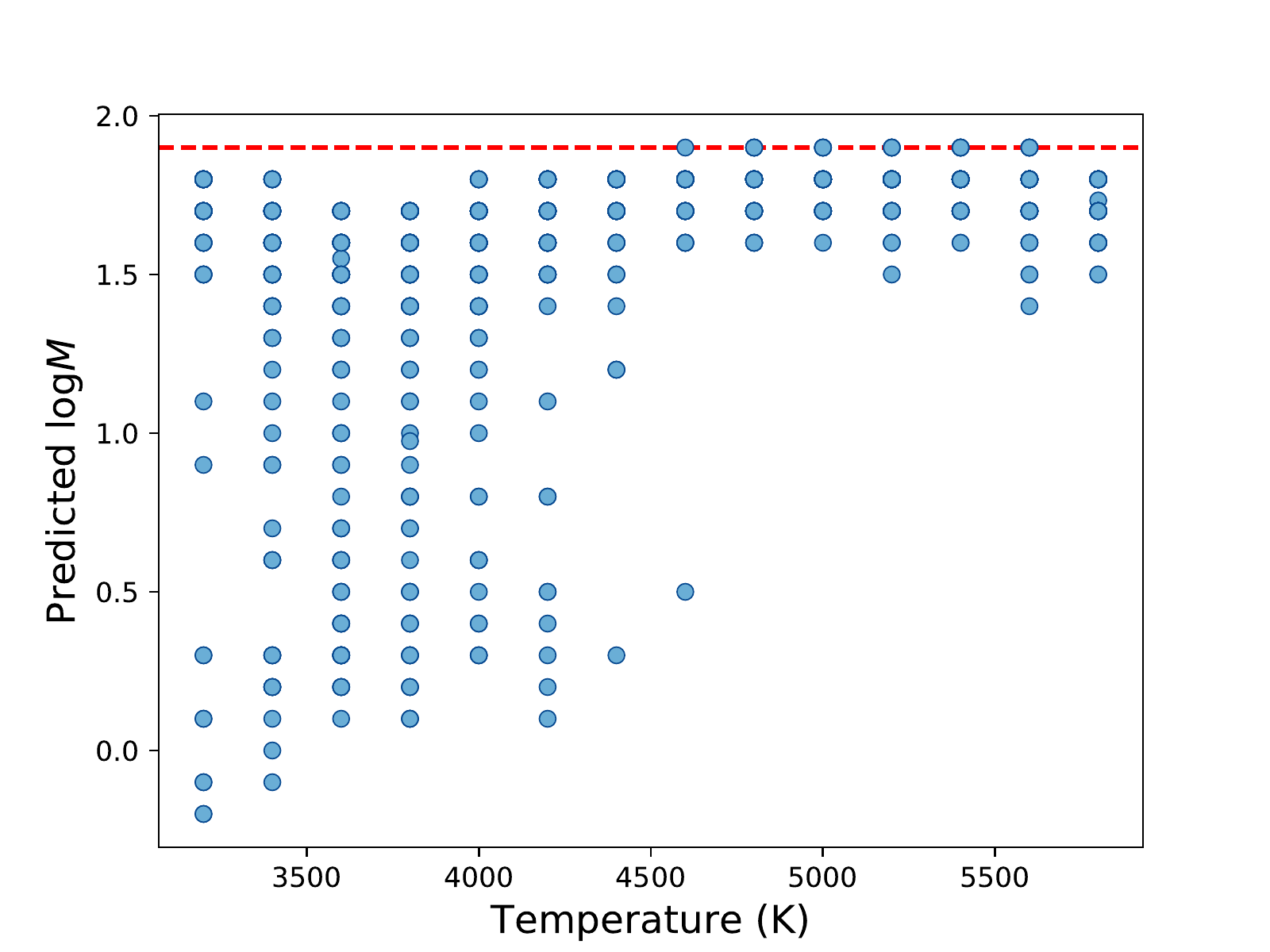}
\end{center}
\caption{Median predictions for metallicity versus the true temperature value for the test spectra with $\log{\rm M}=1.9$, from Figure \ref{fig:ccf_predvreal}. The red, dashed line shows the true metallicity value, 1.9.}
\label{fig:high_met_preds}
\end{figure}

From our tests on the random forest in Figure \ref{fig:ccf_predvreal}, we can see that some of the high metallicity spectra yield much lower metallicity predictions. This is demonstrated further in Figure \ref{fig:met_degen_mock}, which shows a retrieval on one of these high metallicity spectra. The double-peaked posterior leads to a mean prediction that is heavily offset from the true value. This multimodal structure is due to a degeneracy between line depth and metal abundance for high metallicity values. As discussed in Section \ref{sec:rf_mocks}, as the metallicity increases to very high levels, the atmosphere is no longer hydrogen-dominated, causing the mean molecular weight to increase significantly. This in turn decreases the scale height and absorption line depths, reminiscent of lower metallicity values. We tested all the spectra with the highest metallicity value in the testing set ($\log{\rm M}=1.9$), and plotted the median predictions in Figure \ref{fig:high_met_preds}. This plot shows that the degeneracy is stronger at lower temperatures. This follows our physical intuition because at lower temperatures the pressure scale height is smaller, thus compressing the features and reducing the spectrum's sensitivity to metallicity. This makes these spectra harder to distinguish from one another for a given SNR. 

\begin{figure*}
\begin{center}
\includegraphics[width=2.4\columnwidth]{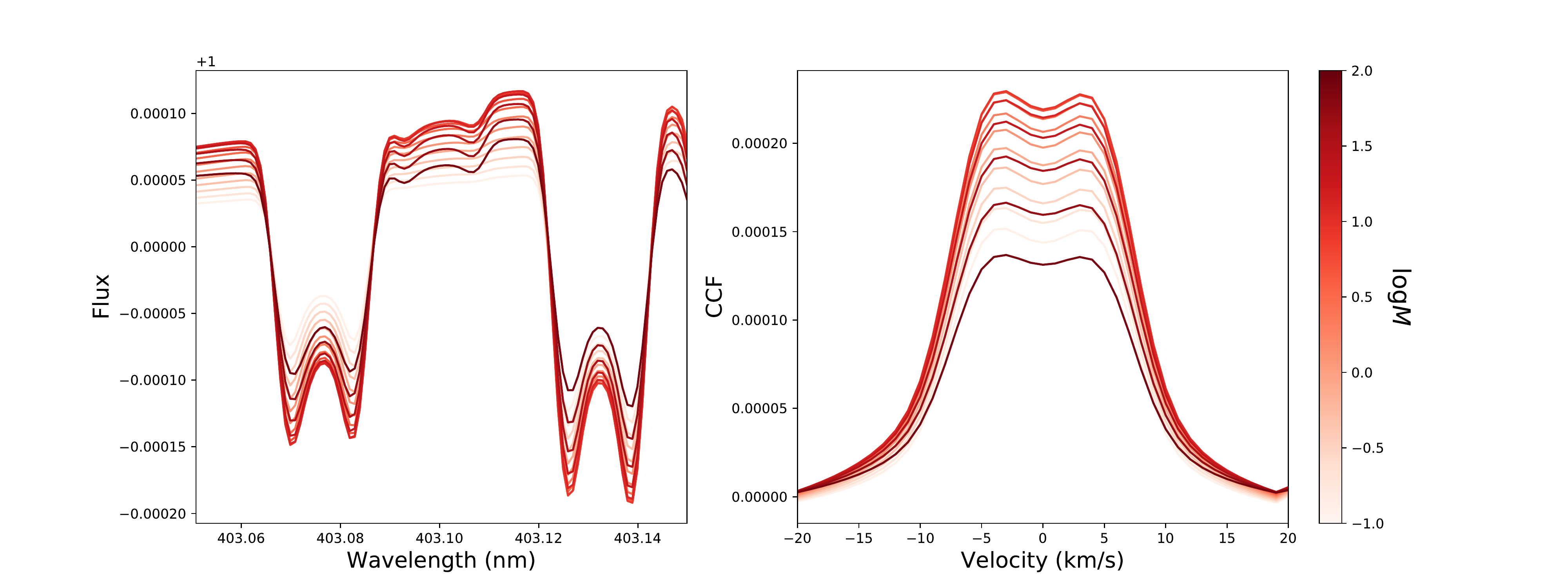}
\end{center}
\caption{Noise-free synthetic spectra with $T=3100$K and varying metallicity values. The left-hand plot shows a zoomed in section of the transmission spectra themselves, whilst the right-hand plot shows a single cross-correlation with each spectrum and the template for Fe at $T=3000$K and $\log{\rm M}=-1.0$. The darker colour corresponds to higher metallicity values.}
\label{fig:met_degen}
\end{figure*}

This degeneracy is also visible in Figure \ref{fig:met_degen}, which shows noise-free spectra with $T=3000$K and varying metallicities, and a cross-correlation with those spectra. As the metallicity increases, the troughs in the left-hand plot deepen up to a point, after which they become shallower again. Similarly, the height of the cross-correlation functions in the right-hand plot increase with metallicity until $\log{\rm M}\gtrsim1.0$, after which the peaks decrease again. Whilst the shape of the high and low metallicity noise-free spectra do differ slightly from each other, these variations are within the error bars of the data, making the noisy spectra indistinguishable. 

\subsection{KELT-9b Retrieval}
\label{sec:kelt9b_retrieval}

\begin{figure}
\begin{center}
\includegraphics[width=\columnwidth]{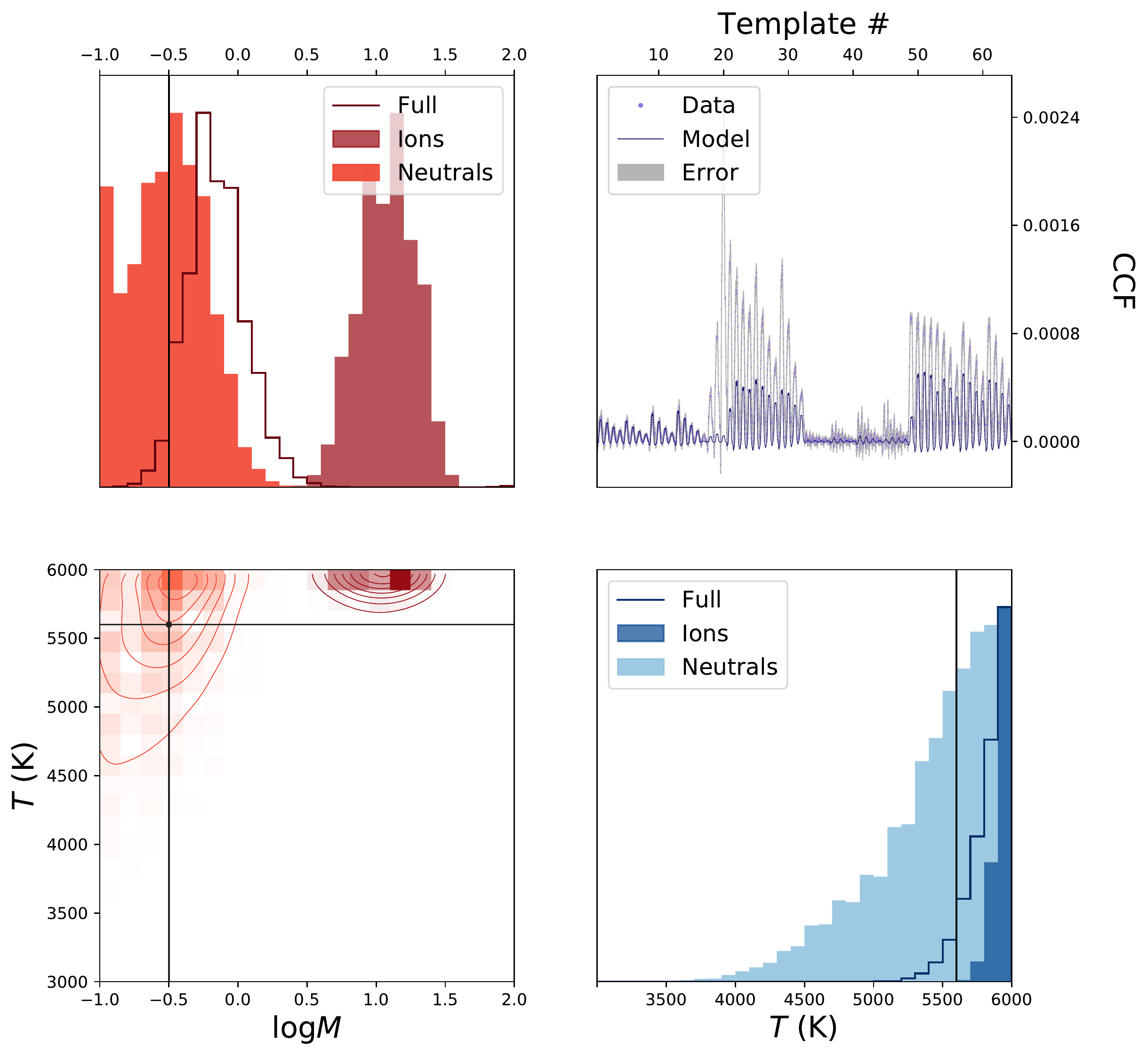}
\end{center}
\caption{Retrieval performed on the CCF-sequence of the transmission spectrum of KELT-9b measured by the HARPS-N spectrograph.  The retrieval is performed in three different ways: using only neutrals (Fe, Ti) (see Figure \ref{fig:neutrals_predvreal}), using only ions (Fe$^+$, Ti$^+$) or using all four species (``Full"). The vertical and horizontal lines indicate the median values of the posterior distributions corresponding to the neutrals-only retrieval. The top right panel shows the data points (lilac) with the error region (grey) for the CCF-sequence produced by the KELT-9b HARPS-N data, along with the model (dark purple) corresponding to the medians from the $\log{M}$ and $T$ posteriors.}
\label{fig:KELT9_retrieval}
\end{figure}

\begin{figure*}
\begin{center}
\hspace*{-0.8cm}
\includegraphics[width=2.4\columnwidth]{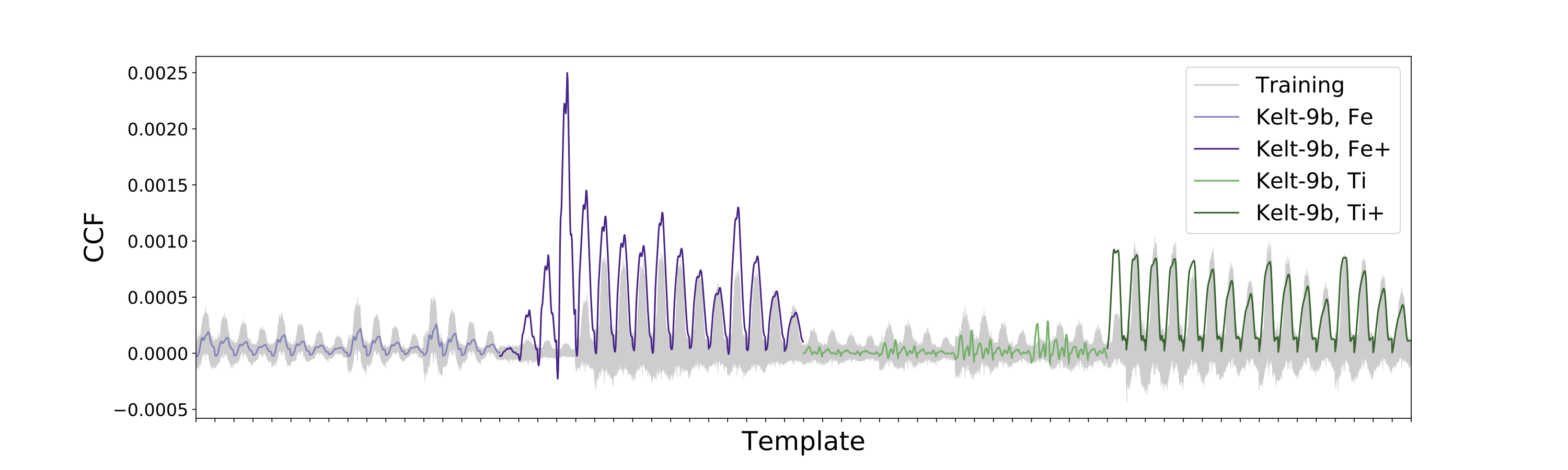}
\end{center}
\caption{Training versus measured KELT-9b CCF-sequences.  The measured CCF-sequence for Fe$^+$ lies outside of the range of the model CCF-sequences, thus flagging missing physics in the model grid.}
\label{fig:KELT9_vs_training}
\end{figure*}

Finally, we performed the hybrid CCF retrieval on the real HARPS-N dataset for the ultra-hot Jupiter KELT-9b. Figure \ref{fig:KELT9_retrieval} shows our results for several different retrievals. As described in \cite{hoeijmakers19}, the ionised iron lines in the spectrum of KELT-9b appear to be much larger than predicted, possibly resulting from an outflowing envelope not present in the model. This leads to a CCF-sequence for the real KELT-9b data that features significantly higher peaks in the Fe$^+$ CCFs when compared to the training set, as shown in Figure \ref{fig:KELT9_vs_training}. With the intent of comparing the effects of the different species, we performed three independent retrievals on the KELT-9b dataset --- one containing the full CCF-sequence, as described in Section \ref{sec:ccf_sequences}, a second containing only the neutral elements, and a third containing only the ions. Each retrieval uses a separate random forest trained on the corresponding sections of the model CCF-sequences. The three retrievals are compared in Figure \ref{fig:KELT9_retrieval}, where the empty lined, darker coloured, and lighter coloured posteriors show the results from the full, ionised and neutral retrievals, respectively. 

The metallicity prediction greatly varies between the different retrievals, which is not unexpected here. The extremely high temperatures cause most of the neutral species to be ionised, leading to low abundances for Fe and Ti. Thus, in the neutral retrieval we predict a low log-metallicity value of $-0.5^{+0.2}_{-0.4}$, whilst the ion retrieval predicts $1.0 \pm 0.2$. The full retrieval lies further towards the neutral prediction, with $\log{\rm M}=-0.2 \pm 0.2$, which is unsurprising due to the stronger feature importance in the neutral CCFs for metallicity. 

\begin{figure}
    \centering
    \includegraphics[width=.93\columnwidth]{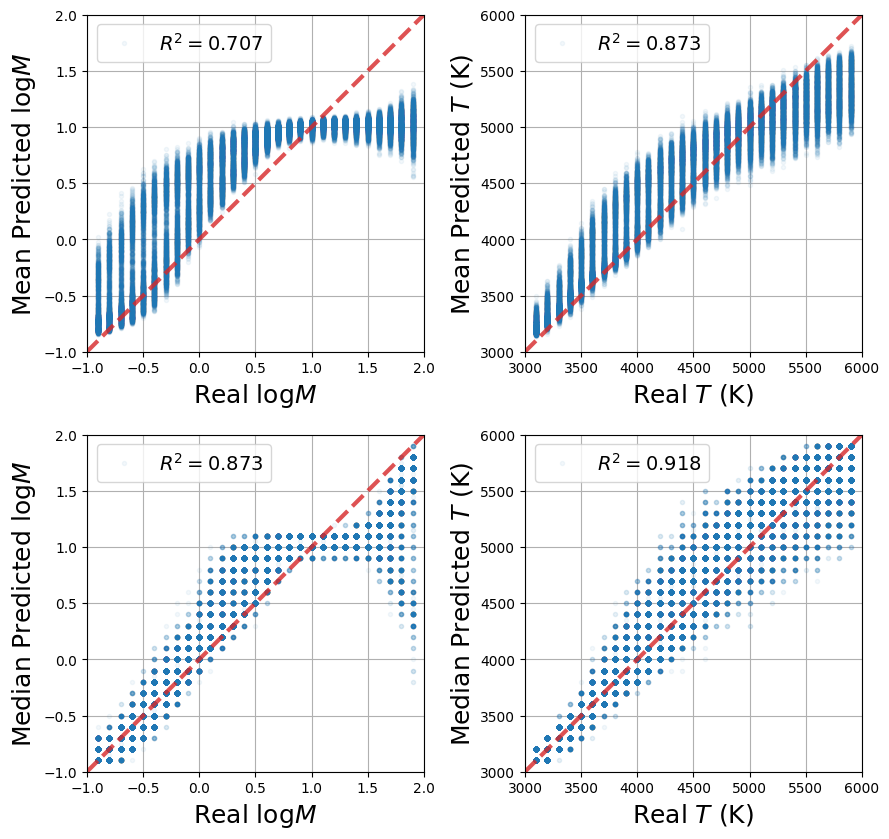}
    \caption{Predicted vs real values for the forest trained on the CCF-sequences with only neutral species, Fe and Ti. The top and bottom panels show the predictions using the means and medians, respectively. The coefficient of determination ($R^2$) varies from -1 to 1, where values near unity indicate strong anti-correlations or correlations between the real and predicted values of a given parameter, based on the variance of the outcomes. The retrieval using this forest on the KELT-9b data is shown as the lighter coloured posteriors in Figure \ref{fig:KELT9_retrieval}.}
    \label{fig:neutrals_predvreal}
\end{figure}

When the Fe$^+$ CCFs are included, i.e. in the full and ion retrievals, the temperature prediction is forced to its upper limit in an attempt to match the strong Fe$^+$ lines ($T=6000^{+0}_{-200}$K and $T=6000^{+0}_{-100}$K for the full and ion retrievals, respectively). However, in the neutral retrieval we still obtain a very high temperature value of $5600^{+400}_{-600}$K, suggesting it is not only the excess Fe$^+$ that escalates the temperature prediction. Figure \ref{fig:neutrals_predvreal} shows the ``predicted vs. real" graphs for the forest trained only on the neutrals. As the temperature increases, this forest's predictive ability decreases, as expected due to ionisation. This suggests that the neutral posterior for temperature in Figure \ref{fig:KELT9_retrieval} may not be reliable. A positive conclusion is that this method is able to identify when a model is flawed. 

Using TESS photometry, \cite{wong19} constrain the dayside and nightside temperatures of KELT-9b to be $4570 \pm 90$K and $3020 \pm 90$K, respectively. However, this is not inconsistent with a higher retrieved temperature from transmission spectroscopy. The dayside spectrum traces higher pressures than the transmission spectrum, which probes tenuous layers of the upper atmosphere. The present retrieval would be consistent with the scenario of an inversion layer, as is predicted in highly irradiated exoplanets \citep{hubeny03,fortney08}.

\section{Discussion}
\label{sec:discussion}

\subsection{Comparison to Nested-Sampling}
\label{sec:nested-sampling}

One of the most common techniques for performing atmospheric retrieval is nested-sampling \citep{skilling06,feroz08,feroz09,feroz13}. In a traditional retrieval, a relatively computationally inexpensive forward model is used to generate spectra on the fly, whilst the sampling method searches the parameter space for the optimal solution. \cite{brogi19} demonstrate a method for performing retrieval on high-resolution data with nested-sampling, but are restricted to $\sim$ 4000 spectral datapoints of the CRIRES instrument. As the number of spectral points increases, so does the time required to compute the models, making this method infeasible for a full HARPS-N spectrum with $\sim$ 300,000 points and multiple free parameters. 

Our method of constructing CCF-sequences allows us to reduce the dimensionality down from $\sim$ 300,000 to $\sim$ 2500. However, now the computational time for each model is much greater as it involves first generating the spectrum and then cross-correlating 64 times with the different templates. Therefore, it remains infeasible to use a standard nested-sampling retrieval for this technique. The random forest requires a grid of pre-computed models to train on, allowing the computational burden to be shifted offline. An alternative method using nested-sampling could be employed by interpolating on the same grid of models, but without the added noise. There are a few disadvantages involved with this when compared with the forest. 

Firstly, the prediction time on a single spectrum is still orders of magnitudes slower than the pre-trained forest ($\sim$ 20 seconds vs $\sim$ 0.05 seconds). This increased computational speed allows the forest to produce ``predicted vs. real" graphs, as shown in Figure \ref{fig:ccf_predvreal} for $\sim$ 50,000 models. These graphs give crucial information about the ability to predict each parameter and the performance of one's retrieval over a vast range of models. We also gain additional information from the random forest, such as the feature importance plots shown in Figure \ref{fig:feat_imp}. This quantifies the information content in each spectral point with respect to each parameter being retrieved, and can be used to infer which areas of the spectrum are most affected by each parameter. It gives us a deeper insight into how the retrieval works, and even indicates which spectral regions might be most informative when considering future observations. 

Secondly, the use of the likelihood function in nested-sampling assumes that the error bars on each spectral point are independent. Whilst this is usually a good assumption, in the process of generating the CCF-sequences we repeatedly cross-correlate a single spectrum with multiple templates, and then concatenate these into a sequence. This implies that the noise samples corresponding to each individual cross-correlation cannot be independent as they propagate from the same spectrum. With this assumption broken, it becomes unclear how to proceed with a nested-sampling retrieval on the CCF-sequences.

\begin{figure}
\begin{center}
\includegraphics[width=\columnwidth]{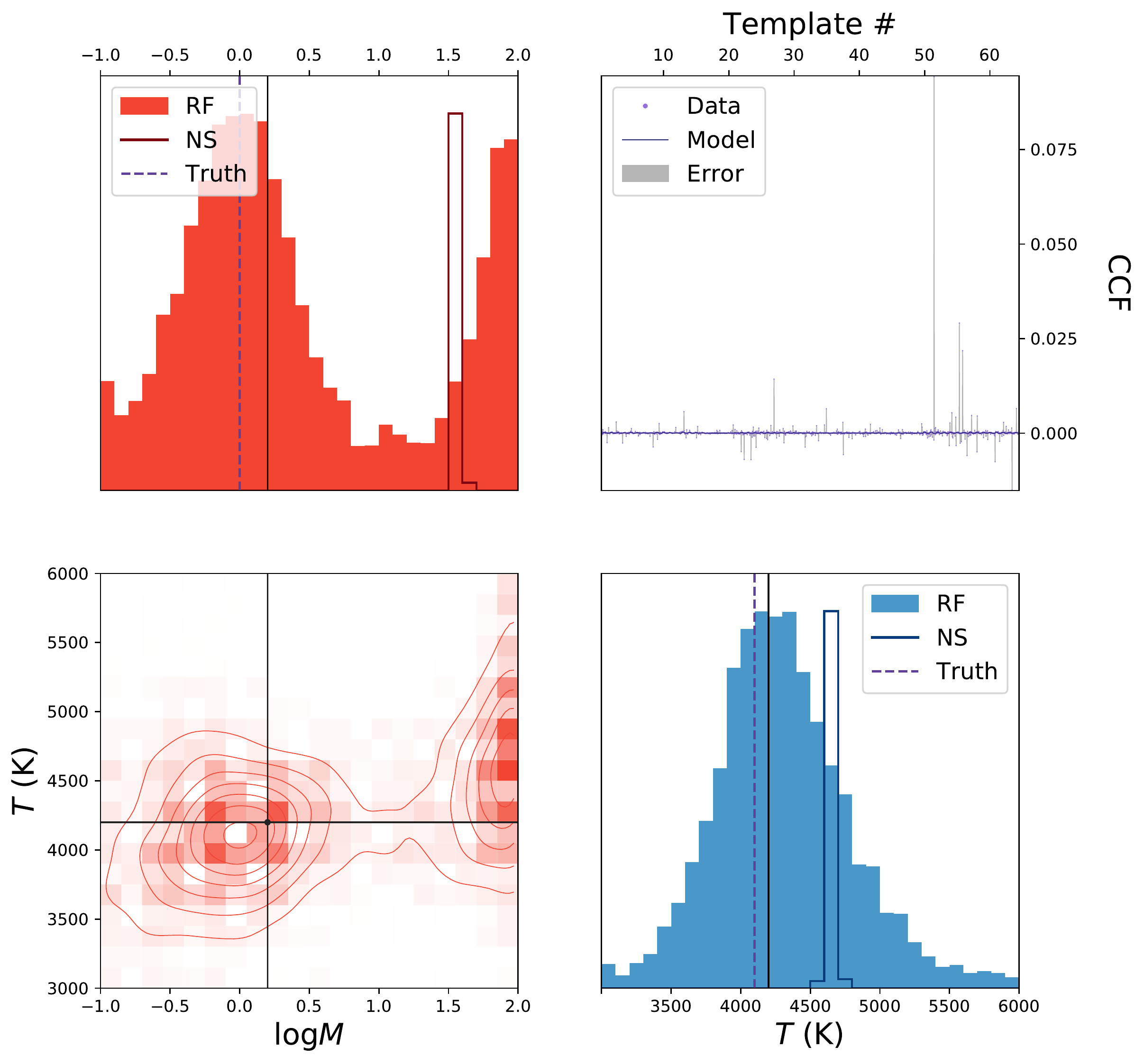}
\end{center}
\caption{A mock retrieval performed on a model with $\log{\rm M}=0$ and $T=4100$K, using the CCF-sequence where the noise has been drawn from a Cauchy distribution. The solid posteriors show the random forest (RF) retrieval results, trained on the CCF-sequences with Gaussian noise models (see Figure \ref{fig:ccf_predvreal}). The empty line posteriors show the nested-sampling (NS) retrieval results using a model that interpolates on the grid of noise-free CCF-sequences and has a Gaussian likelihood. The black lines show the median values of the random forest. The purple, dashed lines show the true values. The top right panel shows the data points (lilac) with the error region (grey), along with the model (dark purple) corresponding to the medians from the $\log{M}$ and $T$ posteriors.}
\label{fig:cauchy_err}
\end{figure}

Thirdly, as discussed in Section \ref{sec:intro_likelihood}, another assumption one needs to make with nested-sampling is a form for the likelihood function, and thus the error bars. For example, it is commonly assumed that the error bars are Gaussians, leading to a likelihood function as shown in equation \ref{eq:likelihood}. The forest also requires an assumption of a random distribution when adding noise to the training set, however it does not depend on a likelihood function. As a test, we generated a model CCF-sequence for a mock retrieval, but this time we added noise by drawing from a Cauchy distribution as opposed to a Gaussian. The motivation behind using a Cauchy distribution is that it does not obey the central limit theorem, and thus the likelihood across many points in a spectrum does not behave as a Gaussian. This provides a challenging test for retrieval methods that assume normally distributed error bars. We performed these retrievals using a forest trained on models with Gaussian errors and a nested-sampling algorithm assuming a Gaussian likelihood function, as shown in equation \ref{eq:likelihood}. Note that this likelihood does not use the cross-correlation function, unlike in \cite{brogi19}. The results are shown in Figure \ref{fig:cauchy_err}. We can see that whilst the posteriors are wide for the forest, they still encapsulate the true values, whereas the nested-sampling retrieval produces tightly constricted, incorrect posteriors. This suggests that the forest is more robust to differences in error distributions.

\subsection{Velocity-Velocity Space Performance}
\label{sec:vv_space}

So far we have only explored the effects of temperature and metallicity, and assumed the velocity parameters, $V_{\rm sys}$ and $K_p$, are fixed to previously determined values. It is possible that neglecting velocities could lead to severe biases in retrievals. To investigate this, we added the systemic velocity ($V_{\rm sys}$) and the error in the semi-amplitude of the planet radial velocity ($\Delta K_p$) to the method. We took $V_{\rm sys}$ from $-10$km/s to $+10$km/s in steps of 2 km/s, and $\Delta K_p$ from 0 km/s to 60 km/s in steps of 6 km/s. Figure \ref{fig:vv_predvreal} shows the results of testing the random forest trained on the CCF-sequences, including the velocity parameters.  

This test shows that the addition of the velocity parameters does somewhat reduce the predictive ability of the other parameters, however this reduction is extremely minor for the temperature, and not too problematic for the metallicity. The method is able to perfectly retrieve the systemic velocity ($V_{\rm sys}$), but struggles with the error in semi-amplitude of the planet radial velocity ($\Delta K_p$). An error in the assumed value of $K_p$ leads to a misalignment of the planet absorption line when summing in the planet rest frame, effectively resulting in a broadening of the CCF. This makes it more challenging to distinguish between sequences of different metallicities. This explains the greater uncertainty in metallicity that we see in Figure \ref{fig:vv_predvreal} when compared with the results from the method without the velocity parameters (Figure \ref{fig:ccf_predvreal}).

\subsection{Comparison to Other Machine-Learning Techniques}
\label{sec:other_ml}

There are several other machine learning methods that can be used to perform atmospheric retrieval \citep{waldmann16,zingales18,cobb19}, each with their own advantages. We tested the same CCF-sequence retrieval as before, but now using a standard neural network and a standard Bayesian neural network (BNN) \citep{gal2016}. In both cases we used a standard multi-layer perceptron architecture with three layers. Each layer consists of a linear transformation with bias followed by a ReLU activation, except the last layer, which does not apply an activation function. The first layer transforms spectra from the input space~$\real^{2560}$ to an intermediate representation~$\real^{512}$. Similarly, layer~2 maps elements to~$\real^{32}$, and layer~$3$ maps elements to the space of parameters~$\real^2$. The Bayesian neural network also applies dropout \citep{srivastava2014} with probability~0.15 on the output of layers~1 and~2, as explained in \cite{gal2016}. We implemented both networks using the \texttt{PyTorch} library for automatic differentiation \citep{paszke2017}, and used \texttt{Adam} \citep{kingma2014} as the optimization method.

The results of the test predictions are shown in Figure \ref{fig:nn_predvreal}. Compared to the random forest, they both perform with slightly improved $R^2$ scores. However, this is only a measure of the average prediction. In atmospheric retrieval, we are predominantly interested in the range of possible parameter values given by a retrieval, and therefore the posteriors of each parameter. A traditional neural network does not produce posteriors, so it cannot be meaningfully applied to this retrieval problem. The BNN does provide posteriors, so we are able to compare these to the forest. Figure \ref{fig:bnn_comp} shows the comparison for two mock retrievals, one with $\log{\rm M}=1.0$ and $T=5100$K (top panel), and one with $\log{\rm M}=1.9$ and $T=4200$K (bottom panel). For the first retrieval, the forest and the BNN produce very similar results, with the BNN posteriors slightly tighter and more centred on the true values. However, in the second retrieval the BNN does not perform well for the metallicity prediction. This mock spectrum was selected as one of retrievals with a strong metallicity degeneracy, as discussed in Section \ref{sec:metallicity_degeneracy}, in order to test how the two methods deal with these issues. The results for the metallicity prediction are $\log{\rm M}=0.3^{+1.7}_{-0.7}$ for the forest, and $\log{\rm M}=0.7^{+0.2}_{-0.2}$ for the BNN. Both the average predictions are heavily offset from the correct value, however the posterior from the forest captures the degenerate behaviour in metallicity, and therefore encompassess the correct value inside the 1-sigma interval. In contrast, the BNN posterior sits in the middle of the degenerate peaks, and remains tightly constrained around the offset value. It is worth noting that this implementation of the BNN is not equivalent to the one used in \cite{cobb19}, as they use a different form of the likelihood which has not been tested on such high-resolution data.

\subsection{Clarification with respect to \cite{cobb19}}

In \cite{cobb19}, it was suggested that the random forest in \cite{marquez18} has the potential to produce over-confident, incorrect posteriors based on a mock retrieval from a test dataset. This forest was trained on WFC3 spectra with 13 data points and predicted 5 parameters --- temperature, free chemical abundances of H$_2$O, HCN and NH$_3$, and a grey cloud opacity, $\kappa_0$. The opacities were calculated with \texttt{HELIOS-K} \citep{grimm15}, using the \texttt{ExoMol}\footnote{http://exomol.com} \citep{tennyson16} spectroscopic linelists for H$_2$O \citep{polyansky18}, HCN \citep{barber14}, and NH$_3$ \citep{yurchenko11}.

The mock spectrum tested on by \cite{cobb19} has $T=1479.6$K, $\log{X_{\rm H_2O}}=-9.79$, $\log{X_{\rm HCN}}=-9.04$, $\log{X_{\rm NH_3}}=-5.91$, and $\log{\kappa_0}=1.87$. The retrieved posterior for NH$_3$ was tightly constrained and offset from the correct value, which was used to infer that the forest could produce spurious results. However, we ran the same retrieval with nested-sampling, using the same model with the same assumptions. Figure \ref{fig:cobb_posterior} shows the results from the random forest retrieval (left panel) and the nested-sampling retrieval (right panel). The posteriors appear very comparable, with the same behaviour in the ammonia abundance.

At the time of publishing \cite{marquez18}, there were no opacity linelists available for NH$_3$ for temperatures above 1500 K. To deal with this, as stated in \cite{marquez18}, the opacity for NH$_3$ was artificially set to zero, and the abundance to the minimum in the range, $10^{-13}$. Notable in this particular mock spectrum is the high cloud opacity, equivalent to a cloud top pressure of $\sim$1 $\mu$bar. This results in an essentially flat spectrum. When retrieving on a flat line, the only two parameters in this model having an effect are the temperature and the cloud opacity, which are perfectly degenerate with each other (i.e. an increase in either results in an upwards shift of the line, so by decreasing the other, one obtains the same spectrum). This degeneracy means one can only obtain lower bounds for the temperature and cloud opacity, corresponding to the upper bound of the other parameter's prior. A consequence of this is a collection of posterior samples in the region $T > 1500$K, which, as forced by the model, have $\log{X_{\rm NH_3}}=-13$, resulting in the peaked posterior for NH$_3$. Therefore, this offset posterior is actually an artefact of the training set, rather than the random forest. This is shown conclusively in Figure \ref{fig:cobb_posterior}, as the forest's posteriors agree with the true Bayesian posteriors from nested-sampling.

\section{Conclusion}
\label{sec:conclusion}

This paper presents a new method for performing atmospheric retrieval on ground-based, high-resolution data of exoplanets. By using a combination of cross-correlation functions we are able to reduce the dimensionality of the problem, and decrease the high levels of uncertainty on each data point. Using our previously demonstrated random forest retrieval technique \citep{marquez18}, we can execute the retrieval quickly and run a multitude of tests of the method. These show that the method performs well on mock data, with a high predictive power for metallicity and temperature ($R^2=0.918$ and $0.986$, respectively). The random forest also provides feature importance plots, which show that the neutral cross-correlations are most important for determining the metallicity, whilst the temperature prediction relies predominantly on the ions. Our method also highlights the metallicity degeneracy in the model, which accounts for the reduced predictability at high metallicity values. 

We performed the retrieval on the HARPS-N data for the ultra-hot Jupiter KELT-9b. The metallicity appears to be consistent with solar, with the retrieval seemingly driven by the neutral species. The prediction for temperature is forced up to exceptionally high values, due to the excess Fe$^{+}$ that appears in the data, suggesting the need for more complex physics in the model. This can be seen when comparing the data to the training set, which also implies that this method is able to recognize when the model is incomplete. 

We also compared the use of our random forest to other approaches, such as the traditional nested-sampling technique and other machine learning methods. We showed that the forest is more robust to the use of different error distributions than nested-sampling, due to it being likelihood-free. When compared with a Bayesian neural network (BNN), although the BNN obtains marginally improved $R^2$ scores, only the forest was able to produce complex posteriors, e.g. in the case of degenerate metallicity values. We also demonstrated that the claim in \cite{cobb19}, that the forest can be over-confident but incorrect, is actually an outcome of the atmospheric model itself and that the forest's posteriors agree with the results from nested-sampling. \\

We thank Heather Knutson for suggesting the Cauchy distribution test, and Ewan Cameron for stimulating discussions on the random forest. We acknowledge financial support from the Swiss National Science Foundation, the European Research Council (via a Consolidator Grant to K.H.; grant No. 771620), the PlanetS National Center of Competence in Research (NCCR), the Center for Space and Habitability (CSH), the Swiss-based MERAC Foundation and the University of Bern International 2021 PhD Fellowship.

\software{\texttt{FastChem} \citep{stock18}, 
        \texttt{HELIOS-K} \citep{grimm15}, 
        \texttt{Helios-o} \citep{bower19}, 
        \texttt{scikit.learn} \citep{pedregosa11}, 
        \texttt{PyTorch} \citep{paszke2017}, 
        \texttt{Adam} \citep{kingma2014}, 
        \texttt{PyMultinest} \citep{buchner14}, 
        \texttt{Astropy} \citep{astropy18}, 
        \texttt{numpy} \citep{vanderwalt11}, 
        \texttt{scipy} \citep{jones01}, 
        \texttt{matplotlib} \citep{hunter07}.
        }

\appendix

\section{Additional Figures}

\begin{figure}[!h]
\begin{center}
\includegraphics[width=\textwidth]{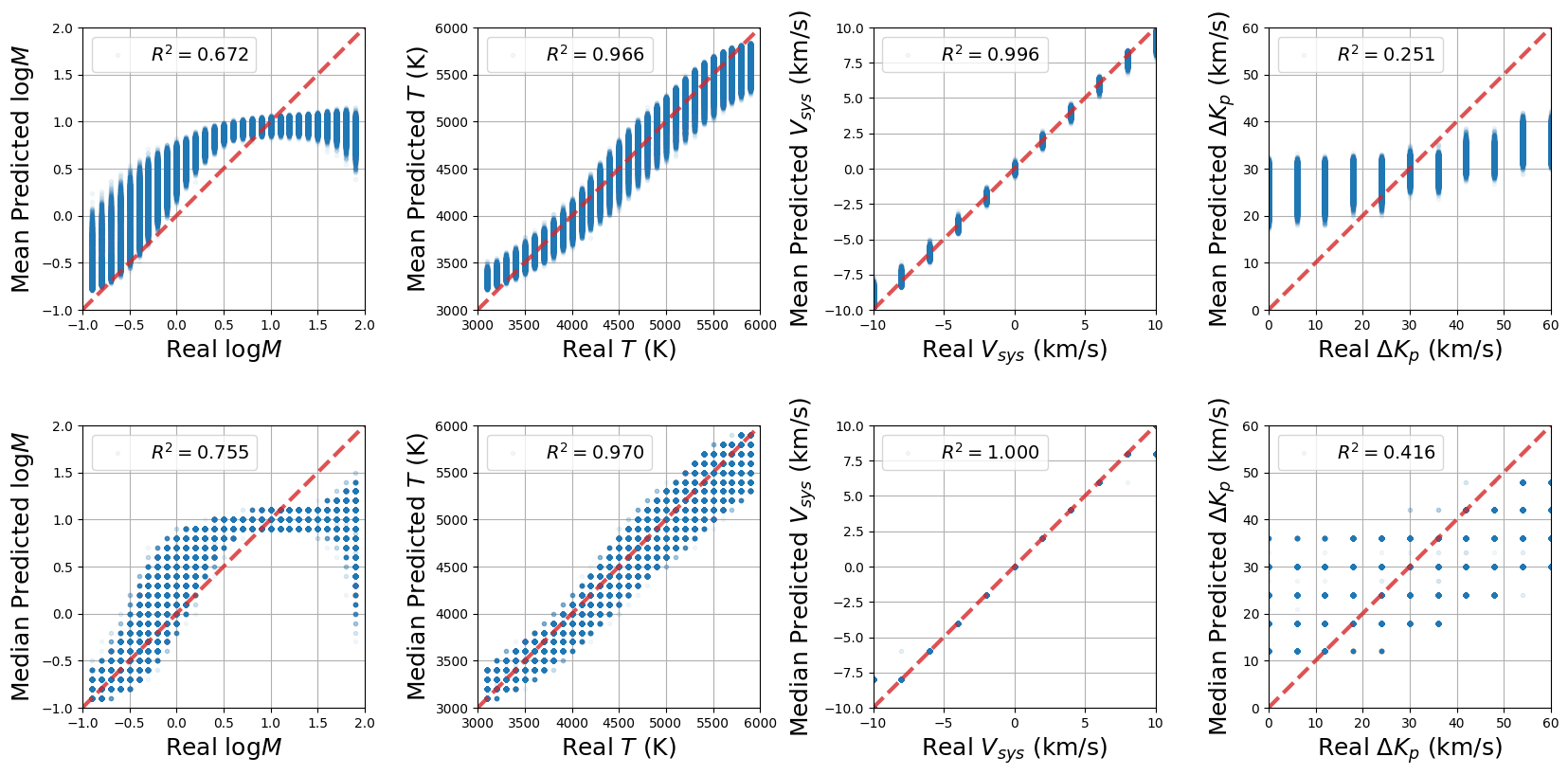}
\end{center}
\caption{Predicted vs real values of the logarithm of metallicity ($\log{\rm M}$), temperature ($T$), systemic velocity ($V_{\rm sys}$) and error in semi-amplitude of the planet radial velocity ($\Delta K_p$) for the random forest trained on the CCF-sequences. The top and bottom rows show the predictions using the means and medians, respectively. The coefficient of determination ($R^2$) varies from -1 to 1, where values near unity indicate strong anti-correlations or correlations between the real and predicted values of a given parameter, based on the variance of the outcomes.}
\label{fig:vv_predvreal}
\end{figure}

\begin{figure}[!h]
\begin{center}
\includegraphics[width=0.49\textwidth]{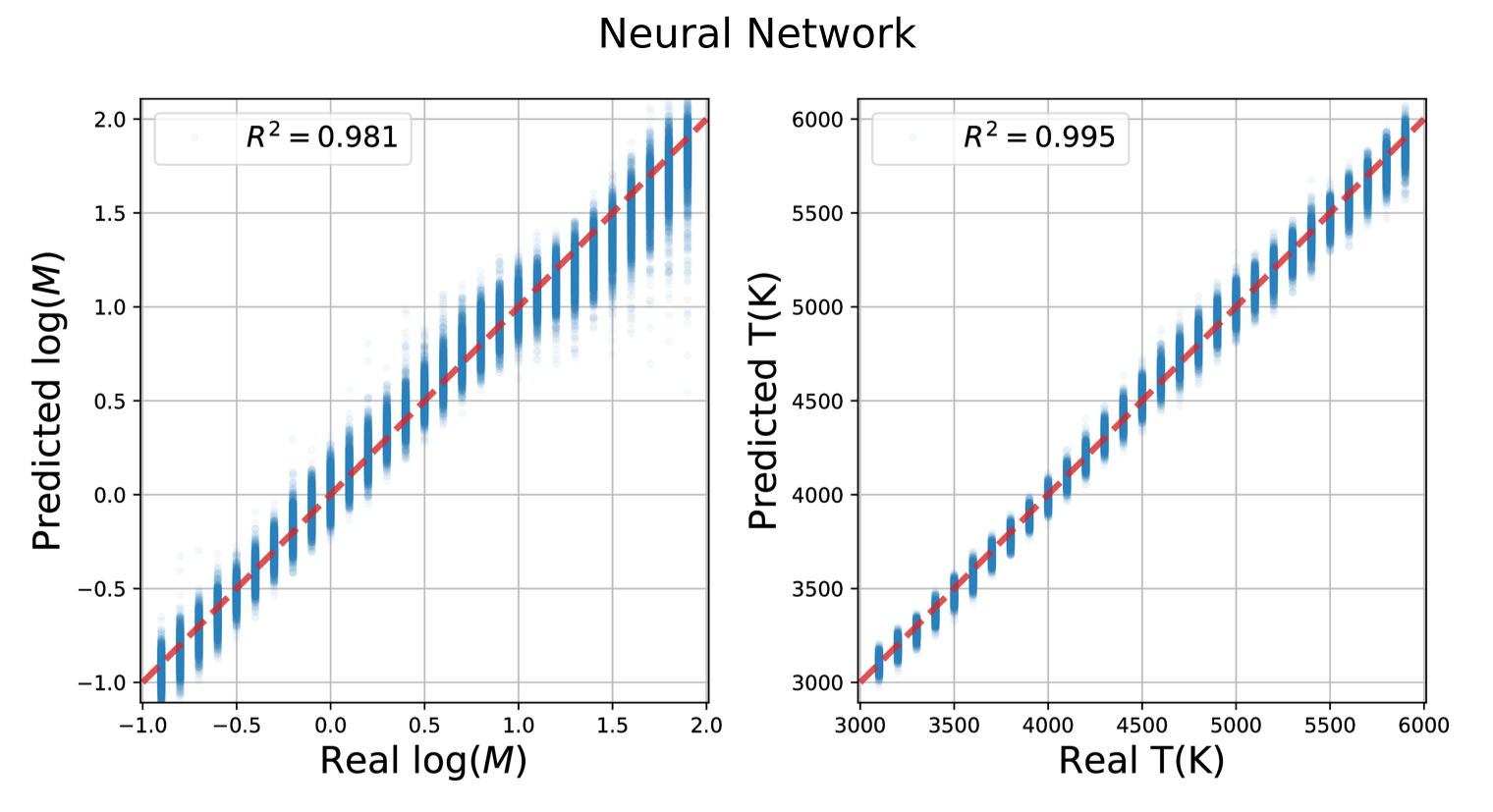}
\hspace{.2cm}
\includegraphics[width=0.49\textwidth]{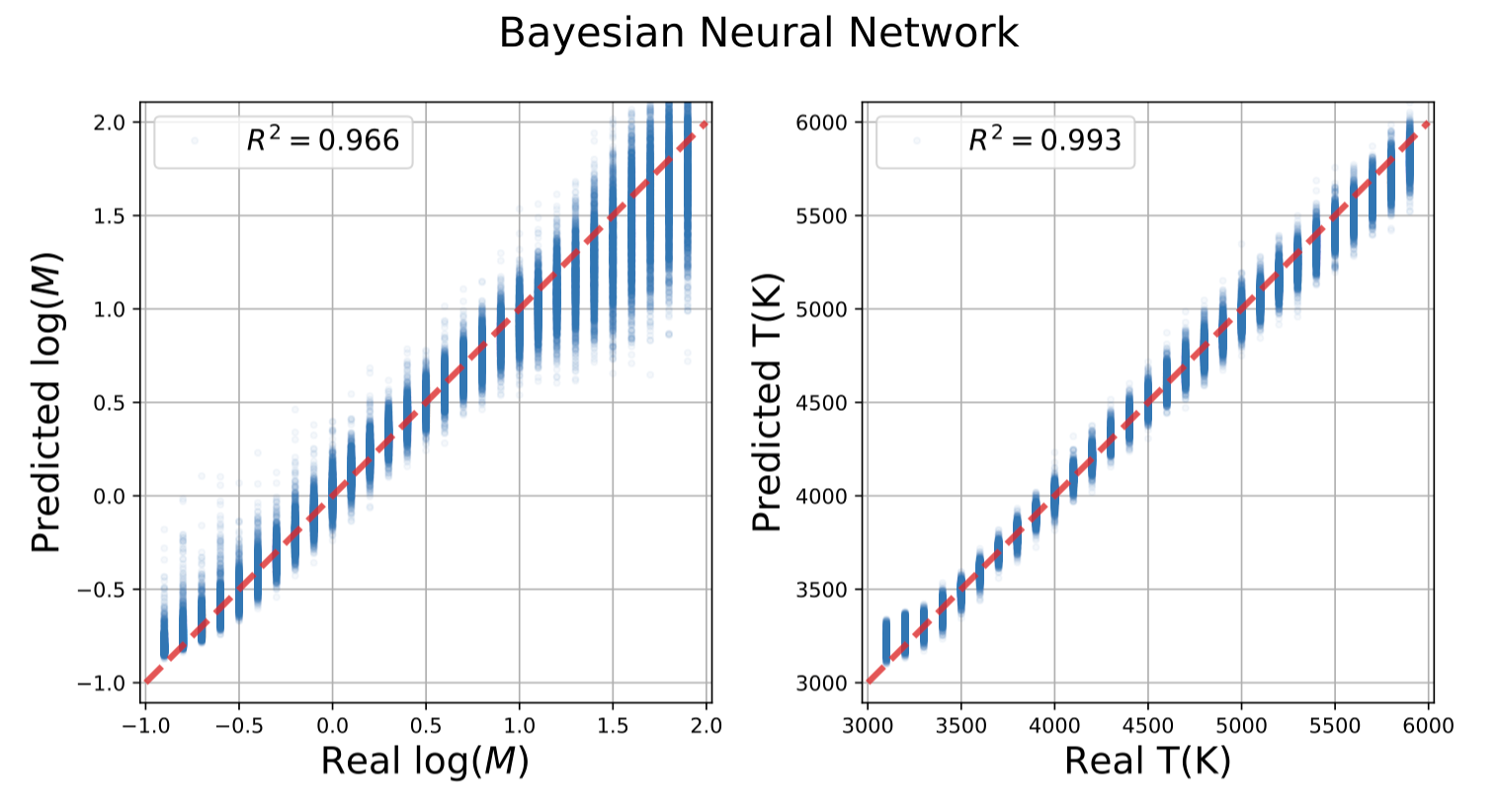}
\end{center}
\caption{Predicted vs. real values for neural networks trained on the CCF-sequences. The left and right pairs show the results for a standard neural network and a Bayesian neural network (BNN), respectively. The coefficient of determination ($R^2$) varies from -1 to 1, where values near unity indicate strong anti-correlations or correlations between the real and predicted values of a given parameter, based on the variance of the outcomes. Mock retrievals for the BNN are shown as the empty line posteriors in Figure \ref{fig:bnn_comp}.}
\label{fig:nn_predvreal}
\end{figure}

\begin{figure}[!h]
\begin{center}
\includegraphics[width=0.48\textwidth]{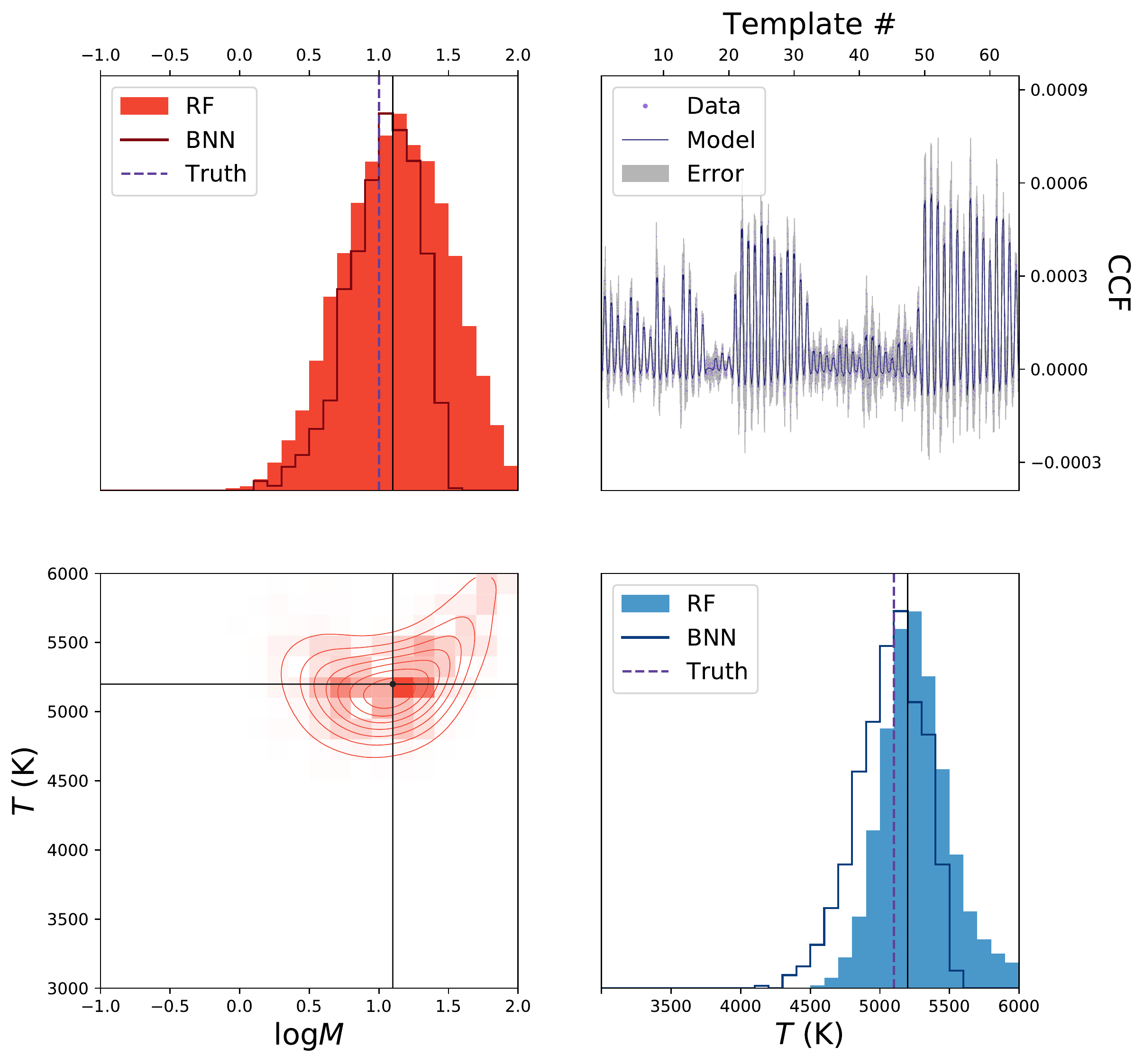}
\hspace{.5cm}
\includegraphics[width=0.48\textwidth]{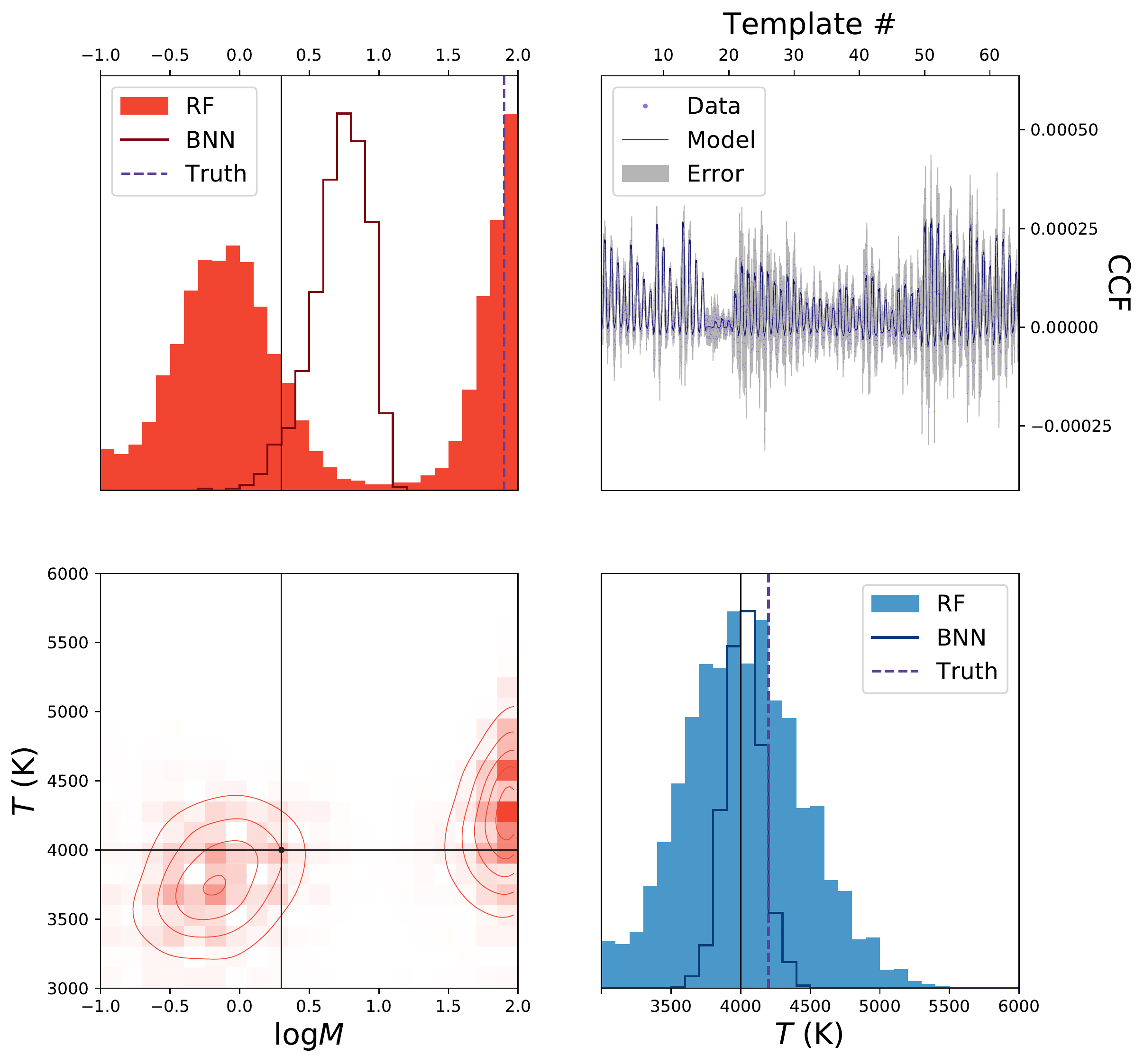}
\end{center}
\caption{Two mock retrievals performed using the CCF-sequences. The solid posteriors show the random forest retrieval results (see Figure \ref{fig:ccf_predvreal}). The empty line posteriors show the Bayesian neural network retrieval results (see bottom panels of Figure \ref{fig:nn_predvreal}). The black lines show the median values of the random forest, and the purple, dashed lines show the true values. The left figure corresponds to a retrieval on a model with $\log{\rm M}=1.0$ and $T=5100$K. The right figure corresponds to a retrieval on a model with $\log{\rm M}=1.9$ and $T=4200$K.}
\label{fig:bnn_comp}
\end{figure}

\begin{figure*}[!h]
\begin{center}
\includegraphics[width=0.48\textwidth]{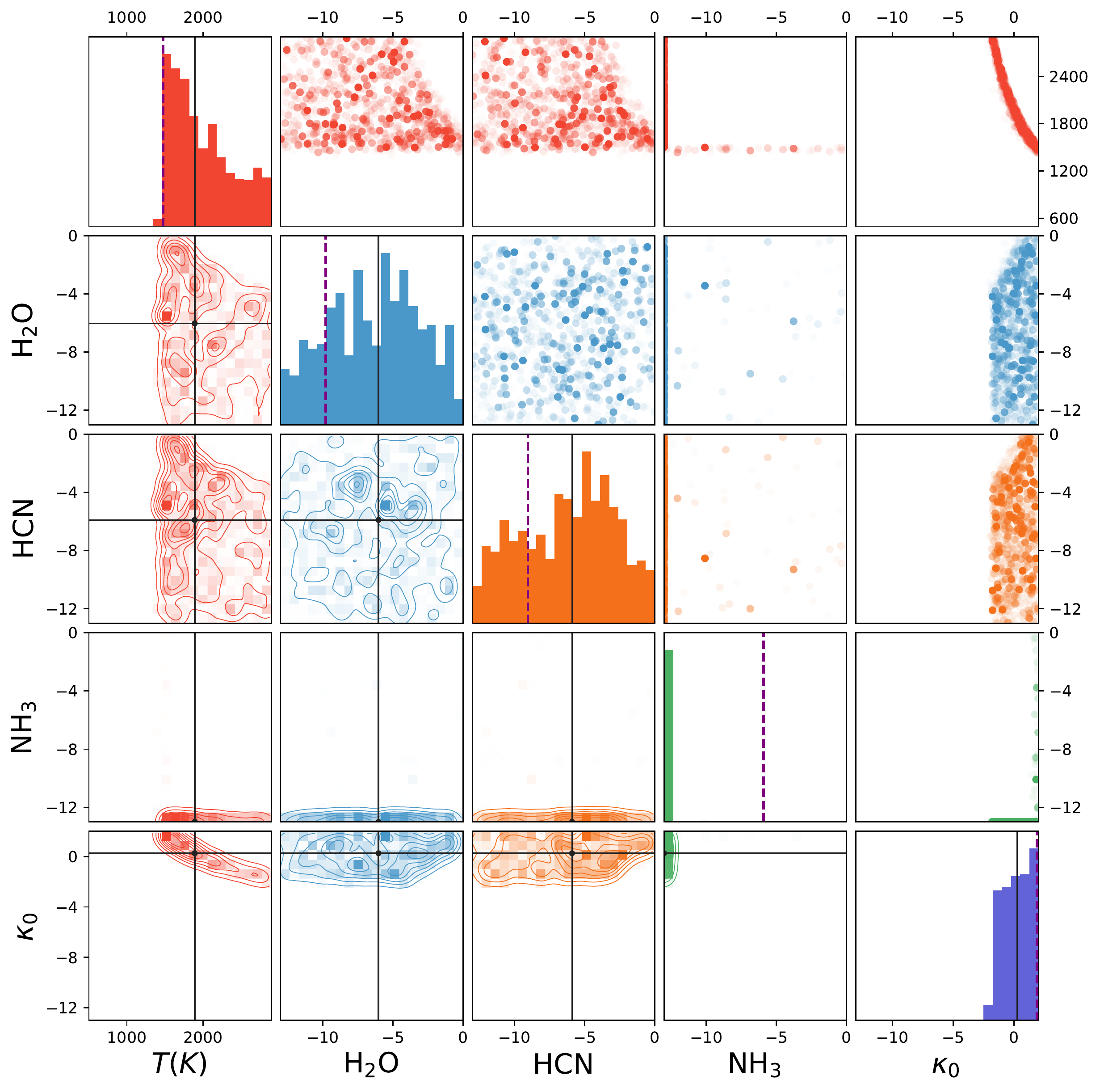}
\hspace{.5cm}
\includegraphics[width=0.48\textwidth]{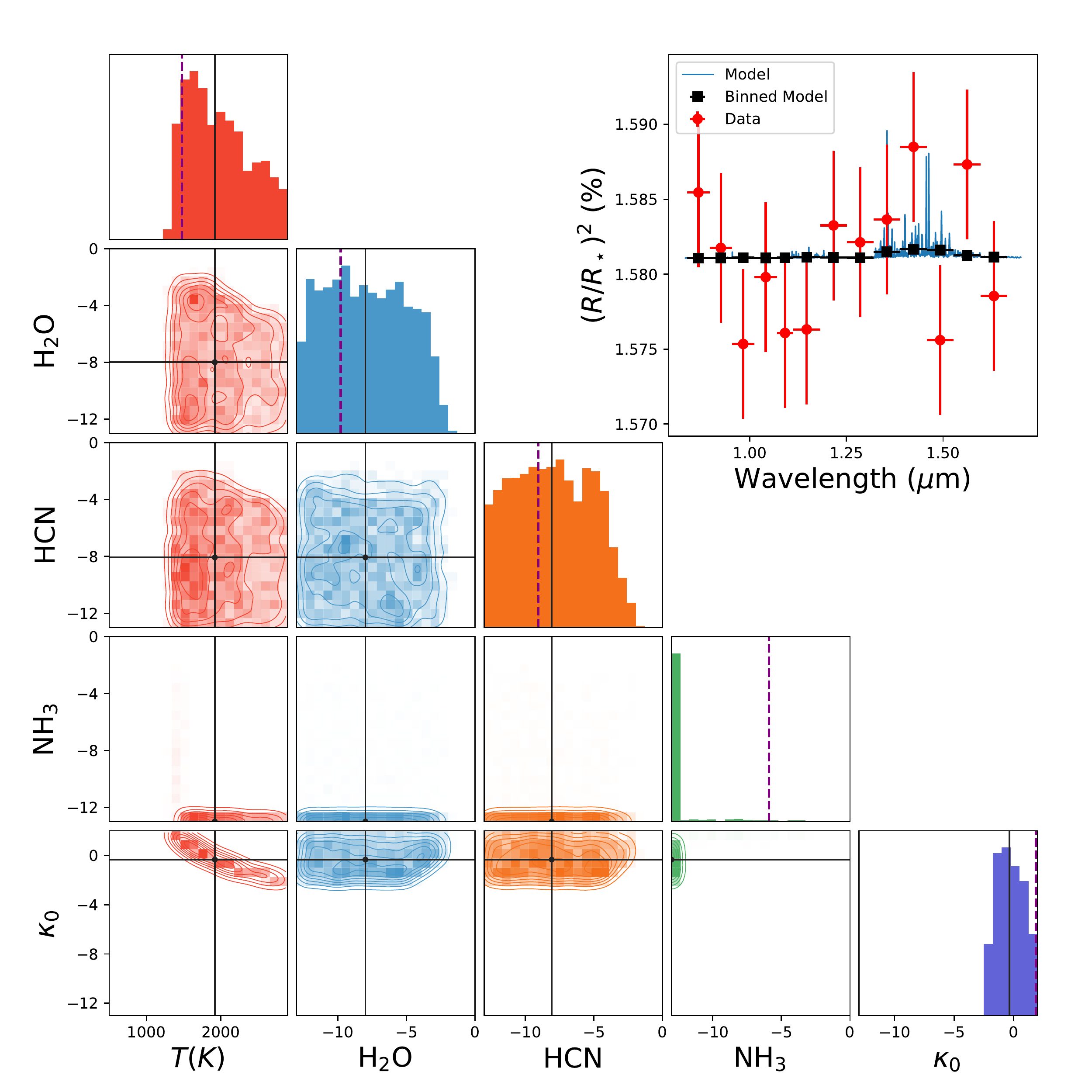}
\end{center}
\caption{Retrieval results for a mock spectrum with $T=1479.6$K, $\log{X_{\rm H_2O}}=-9.79$, $\log{X_{\rm HCN}}=-9.04$, $\log{X_{\rm NH_3}}=-5.91$, and $\log{\kappa_0}=1.87$. The left- and right-hand plots show the results using a random forest and nested-sampling, respectively. The black lines show the median predictions. The purple, dashed lines show the true values.}
\label{fig:cobb_posterior}
\end{figure*}


\label{lastpage}

\end{document}